\documentclass[aps,pre,reprint]{revtex4-2}
\usepackage{amsmath}
\usepackage{bm}

\newcommand{\mb}{\mathbf}   
\newcommand{\Ev}[2]{\left.{#1}\right|_{#2}}
\newcommand{\dd}[2]{\frac{\partial #1}{\partial #2}}
\providecommand{\DD}[2]{\frac{\D #1}{\D #2}}
\newcommand{\TwoVect}[2]{\bmatrix #1 \\ #2 \endbmatrix}
\newcommand{\abs}[1]{\left|{#1}\right|}
\newcommand{\rD}{\partial}

\newcommand{\dt}{\partial_t}

\newcommand{\gm}{\gamma}

\newcommand{\ld}{\lambda}

\newcommand{\psiD}[1][\relax]{{#1\psi}_{\mathrm{d}}}
\newcommand{\psiR}[1][\relax]{{#1\psi}_{\mathrm{r}}}
\newcommand{\PsiD}[1][\relax]{{#1\Psi}_{\mathrm{d}}}
\newcommand{\PsiR}[1][\relax]{{#1\Psi}_{\mathrm{r}}}

\newcommand{\Cd}{{C}_{\mathrm{d}}}
\newcommand{\Cr}{{C}_{\mathrm{r}}}
\newcommand{\Id}{\tensor{I}_{\mathrm{d}}}
\newcommand{\Ir}{\tensor{I}_{\mathrm{r}}}

\newcommand{\Av}[1]{\left\langle{#1}\right\rangle}
\newcommand{\kB}{k_{\text{B}}}
\newcommand{\kT}{{\kB}T}

{}%
\newcommand{\BoldXi}{\bm{\xi}}

\newcommand{\Dc}{D^{\text{c}}}
\newcommand{\Fs}{F_{\text{S}}}

\newcommand{\ChiR}{\mathsf{X}}
\newcommand{\para}{\parallel}
\newcommand{\Xl}{X_\para}
\newcommand{\Xtr}{X_\perp}
\newcommand{\di}[1][\mb]{\tilde{#1{d}}}
\newcommand{\fd}{\check{f}_\text{d}}
\newcommand{\fr}{\check{f}_\text{r}}

\newcommand{\muR}{{\mu_{\text{r}}}}
\newcommand{\D}{\mathrm{d}}

\newcommand{\Vmax}{V_{\text{max}}}

\DeclareMathOperator{\erfc}{erfc}
\usepackage{url}
\newcommand{\Lemaitre}{Le\-ma\^\i\-tre}
\newcommand{\br }{\notag\\&\quad{}}
\newcommand{\brX}{\\{}}
\newcommand{\hFil}{\relax}

\usepackage{graphicx}
\graphicspath{{Fig/}{../Fig/}{.}}
\usepackage{color}
\newcommand{\IK}{\mathrm{i}\mb{k}}
\newcommand{\Ik}{\mathrm{i}{k}}
\newcommand{\phiA}{\phi_{\text{area}}}
\newcommand{\tauA}{\tau_\alpha}

\newcommand{\K}{\mb{k}}

\newcommand{\XiR}{\BoldXi_*}
\newcommand{\dXiR}{\dd{}{\xi_*}}
\newcommand{\dPhi}{\dd{}{\varphi_*}}
\newcommand{\tD}{t_\Delta}
\newcommand{\Sd}{S_{\mathrm{d}}}
\newcommand{\Sr}{S_{\mathrm{r}}}
\newcommand{\CosPh}{\cos{\varphi_*}}
\newcommand{\CosSqPh}{{\cos^2}{\varphi_*}}
\newcommand{\SinPh}{\sin{\varphi_*}}
\newcommand{\SinSqPh}{{\sin^2}{\varphi_*}}
\newcommand{\dXi}{\partial_1}
\newcommand{\dXiPrime}{\partial_{1'}}
\newcommand{\dXiAst}{\partial_{1_*}}
\newcommand{\dHa}{\partial_2}
\newcommand{\dHaPrime}{\partial_{2'}}
\newcommand{\dHaAst}{\partial_{2_*}}
\newcommand{\Order}{\mathcal{O}}
\newcommand{\de}{\bar\delta_\varepsilon}
\newcommand{\Ix}{\alpha}
\newcommand{\gmEM}{\gamma_{\text{EM}}}
\newcommand{\kmax}{k_{\text{max}}}

\begin{document}
\title{Displacement correlations 
       in a two-dimensional colloidal liquid
       and their relationship with shear strain correlations}
\author{\surname{Ooshida} Takeshi}
\email[E-mail:~]{ooshida@tottori-u.ac.jp}
\affiliation{%
   Department of Mechanical and Physical Engineering,
   Tottori University, Tottori 680-8552, Japan}
\author{Takeshi \surname{Matsumoto}}
\affiliation{%
   Division of Physics and Astronomy, Graduate School of Science, 
   Kyoto University, 
   \relax{Kyoto 606-8502, Japan}}
\author{Michio \surname{Otsuki}}
\affiliation{%
   Graduate School of Engineering Science,  
   Osaka University, 
   \relax{Toyonaka, Osaka 560-8531, Japan}}

\thispagestyle{empty}%
\begin{abstract}
Correlations of the displacement field 
  in a two-dimensional model colloidal liquid 
  is studied numerically and analytically.
By calculating 
  the displacement correlations 
  and the shear strain correlations
  from the numerical data of particle simulations, 
  the displacement field is shown to have nontrivial correlations, 
  even in liquids that are only slightly glassy
 with the area fraction as low as ${0.5}$.
It is suggested analytically and demonstrated numerically 
  that the displacement correlations 
  are more informative than the shear correlations:
  the former behaves logarithmically 
  with regard to the spatial distance at shorter scales, 
  while the corresponding information 
  is missing from the shear correlations.
The logarithmic behavior of the displacement correlations
  is interpreted as manifesting 
  a long-lived aspect of the cage effect. 
\end{abstract}%
\maketitle
\section{Introduction}
\label{sec:intro}

Dense liquids are sometimes described 
  as ``solids that flow'' \cite{Nakagawa.Book1975jx,Dyre.PRE76}.
This description,
  not only macroscopically but also microscopically, 
  offers an appealing point of view,  
  complementary to the kinetic approach to liquid states.
The kinetic approach \cite{Balucani.Book1994}
  is an extension of the molecular theory of gases, 
  in which displacements 
  do not play an essential role.
In the case of dilute fluids such as gases,
   the molecular velocities are totally different 
  from the hydrodynamic velocity field.
In such a case, 
  it would be a tremendous mistake 
  to expect that the time-integral of the hydrodynamic velocity field 
  gives the displacements of the molecule, 
  as if the so-called ``fluid parcels'' 
  were really wrapped by a membrane that confines the molecules.  
  
For liquids of extremely high density, however, 
  it is justifiable to define the displacement field 
  by time-integral of the hydrodynamic velocity field \cite{Klix.PRL109}.
As long as the liquid particles (either molecular or colloidal) 
  are basically confined in cages, 
  the displacement of each particle 
  can be approximated 
  by the value of the displacement field at its position. 
This is a feature 
  in common with solids.
As is well known in the case of crystalline solids, 
  spatially correlated displacements 
  are associated with elastic stresses. 
It is therefore natural to expect 
  that, also in the cases of such high-density liquids 
  close to vitrification (often described as ``glassy''), 
  the correlations of the displacement field  
  may provide information about elasticity.

An interesting question 
  is whether the displacements are correlated 
  also in just slightly glassy liquids.
Before stating this question more precisely, 
  we need to take notice of various statistical quantities 
  that have been devised 
  by using the displacement as a main ingredient.
Some of them are based on scalar quantities
  such as the density of mobile particles, 
  while others are grounded on tensorial quantities
  possibly related to elasticity.

In numerical studies 
  resolving the particle displacements in both space and time, 
  researchers have noticed 
  that mobile particles are distributed heterogeneously 
  \cite{Yamamoto.PRE58,Berthier.RMP83}.
The length scale of this dynamical heterogeneity 
  has been studied 
  typically through correlation of some quantity 
  intended to represent the density of mobile or immobile particles, 
  such as 
  the magnitude of the particle displacement 
  \cite{Donati.PRL82}, 
  the overlap density \cite{Glotzer.JCP112}, 
  and the self part of the intermediate scattering function 
  \cite{Berthier.PRE69,Toninelli.PRE71}.

The formulations mentioned above 
  are focused on the mobility of particles
  and not on their direction of motion. 
The importance of the directional aspect of motion 
  was noticed by Doliwa and Heuer \cite{Doliwa.PRE61}, 
  who demonstrated with Monte Carlo simulations 
  that the displacement correlations in the longitudinal direction 
  behave quite differently from those in the transverse direction.
Essentially the same spatial pattern of displacement correlations 
  was also observed independently 
  by Cui \textit{et al.}~\cite{Cui.PRL92} 
  in experiments with aqueous suspension of silica spheres.
Later, it was noticed 
  that the transverse displacement correlations 
  contain information about glass elasticity  
  \cite{Klix.PRL109,Flenner.PRL114,Ikeda.PRE92}.
Thus we are interested 
  in tensorial correlations of displacements in glassy liquids, 
  which provides a key topic of the present work.  

While the displacement in solids  
  in purely elastic response to external loading 
  has basically smooth dependence on the spatial coordinate 
  even at meso\-scopic scales,
  its inelastic responses can be more localized, 
  as is exemplified by motion of dislocations in crystalline solids.
In the case of glassy materials 
  subject to external shear, 
  inelastic deformation occurs 
  in the form of a localized plastic event, 
  which, in turn, 
  induces deformation of the surrounding elastic medium
  \cite{Picard.EPJE15,Adhikari.arXiv2306}.
The response of elastic media 
  to such a localized inelastic source  
  is given by the Eshelby strains 
  \cite{Eshelby.RSPA241}.
Features of the Eshelby strains
  were reproduced from data of particle displacements 
  by calculating correlations of meso\-scopic shear strain field
  \cite{Chikkadi.PRL107,Chattoraj.PRL111,Illing.PRL117}, 
  which has been taken as the signature of elasticity.
The Eshelby strain pattern observed with this method,
  interestingly,   
  persists 
  not only in sheared glasses behaving as elasto\-plastic materials
  but also in glassy liquids, with or without shear, 
  even on timescales longer than the structural relaxation time.
  
Motivated by the above considerations, 
  here we concretize our interest 
  in displacement-based statistical quantities
  by discussing  
  \emph{two-particle displacement correlations} (DC)  
  and 
  \emph{shear strain correlations} (SC)
  in two-dimensional liquids.
As is exemplified by the aforementioned experiment 
  by Cui \textit{et al.}~\cite{Cui.PRL92},
  DC and SC can be measured experimentally.
Measurement of DC between tracer particles in viscoelastic media 
  has been applied to various soft materials  
  by the name of ``two-point micro rheology'' 
  \cite{Crocker.PRL85,Mizuno.SoftMatter16}.
DCs of directly interacting colloidal particles 
  in a suspension with finite density 
  have also been reported \cite{Dell.PRE92}.
Besides, 
  Illing \textit{et al.}~\cite{Illing.PRL117} 
  conducted video microscopy experiments on SC in a glassy liquid.
To connect these experimentally (and computationally) measurable 
  statistical quantities 
  with the idea of solid-like liquid theory, 
  now we raise three questions about DC and SC.

Our first question 
  is whether and to what extent 
  these correlations, namely DC and SC, are detectable 
  in liquids that are only slightly dense. 
To demonstrate the presence of these correlations in liquids
  is to legitimate the solid-based approach to the liquid dynamics.
We focus on such density
  that the lifetime of ``bonds'' between the nearest neighbors 
  is barely non-zero and not very long. 
Although the liquid in such cases 
  does not exhibits macroscopic elasticity at all, 
  we will show numerically 
  an appreciable presence of DC and SC 
  at length\-scales greater than the particle diameter.
The time dependence of these correlations 
  will also be investigated.

To make better use of DC and SC 
  as experimentally measurable quantities, 
  we ask further questions about their nature.
The second question 
  concerns the possible equivalence between the DC and the SC, 
  as both of them are calculable 
  from the same data of particle displacements.
Does the usage of the common ingredient 
  mean that the two kinds of correlations are equivalent?
We will answer this question negatively, 
  showing that the SC can be calculated from the DC 
  but not vice versa.  
In other words, 
  the DCs are more informative than the SCs, 
  which leads to the third question: 
  what is the nature 
  of the extra information contained in DC?
This question will be answered, at least partially, 
  by showing that the DCs behave logarithmically for shorter distances,   
  with the coefficient of the logarithmic term 
  being an increasing function of the density.
We interpret the nature of this extra information
          as manifestation of the cage effect.


The paper is organized 
  as follows.
We will start by reviewing some background about DC and SC 
  and giving their definitions in Sec.~\ref{sec:BG}. 
Subsequently, in Sec.~\ref{sec:relation}, 
  we will develop a theoretical framework 
  for analytical treatment of SC;
  the framework,
  which is constructed by extending 
  a previous work of our group on DC \cite{Ooshida.PRE94},
  allows us to express SC in terms of DC.
After specifying the two-dimensional Brownian particle system
  in Sec.~\ref{sec:system} as a model liquid,
  we will compute the DC and the SC 
  from the numerical data in Sec.~\ref{sec:num}.
As a result, 
  the DCs will be shown to behave logarithmically 
  at length\-scales 
  between the particle diameter and the correlation length, 
  which makes the DC more sensitive to the mean density 
  (measured by the area fraction) 
  than the SC.
Possible origin of the logarithmic behavior of DC 
  is discussed in Sec.~\ref{sec:discussion},
  along with the fact 
  that the DC and the SC can be expressed 
  in terms of a similarity variable with diffusive dynamical length.
Finally, 
  Sec.~\ref{sec:conc} is devoted to concluding remarks.

\section{Background}
\label{sec:BG}

\subsection{Caged dynamics in liquids as ``solids that flow''}
\label{subsec:liquids}
  
As a background to the questions about DCs and SCs in liquids, 
  we begin this review section 
  with the notion of cage effects in dense liquids  
  regarded as solids with impermanent bonds.

We have noted in Introduction 
  that the approach from the solid-like aspect of liquids 
  is complementary to that by extending
  the kinetic theory of gases toward higher densities. 
The kinetic approach \cite{Balucani.Book1994}
  seems more naturally applicable 
  when the correlations between collisions are smaller;
  this occurs if the density is so low 
  that collision is basically a process 
  in which a molecule encounters a new partner every time.  
In other words, reunion with the same molecule 
  (as well as ring collision) 
  is a rare event in dilute fluids. 

In dense liquids, contrastively, 
  neighboring particles stay neighbors 
  for a long time. 
This occurs 
  even in the absence of attractive interactions.
Every particle tends to remain 
  in contact with the same neighbors,
  often described as a cage 
  in which the particle is trapped.
Although the notion of repeated collisions 
  is still applicable 
  to the interaction between the caged particle and its neighbors, 
  the effect of ceaseless interaction
  between the same pair of particles 
  is more appropriately regarded as a kind of bond 
  with a long but finite lifetime
  \cite{Yamamoto.PRE58,Shiba.PRE86}.
In the limiting case of permanent bonds, 
  the material would behave 
  as a solid (either crystalline or amorphous).

The presence of ``bond breakage'' as a rare event 
  introduces fluidity.
The lifetime of bonds, in the absence of external shear,  
  is on the order of the structural relaxation time $\tauA$,
  conventionally defined 
  as the timescale of the intermediate scattering function 
  at the scale of the particle diameter \cite{Kob.PRE52}.

Thus the microscopic dynamics of dense liquids
  for time\-scales shorter than $\tauA$
  are solid-like, 
  in the sense 
  that they are dominated by such ``bonds'' with a long lifetime.
The total effect of such a bond structure 
  could be conceived as confinement by nested cages, 
  and therefore often referred to as the cage effect.
For time\-scales longer than $\tauA$, 
  the ``bonds'' are broken 
  and therefore the dynamics cannot be purely elastic,
  but we note the possibility 
  that the collapse of the whole cage structure 
  may take much longer time than $\tauA$. 
The dynamics are then expected to be viscoelastic, 
  possibly depending on length\-scales 
  in some intriguing manner.

\subsection{Displacement in liquids}
\label{subsec:displacement}

Let us continue reviewing 
  statistical physics of dense liquids in general, 
  which can be molecular or colloidal for the present.

The cage effect
  suggests the possibility 
  that, in dense liquids, hydrodynamic description may be extended 
  to shorter length\-scales close to the inter\-particle distance
  \cite{Balucani.Book1994,Mountain.AMRP9}. 
Hydrodynamics means 
  finding a closed set of dynamical equations for slow variables. 
The long lifetime of the ``bonds'' 
  can provide grounds for slow variables in dense liquids, 
  such as the \emph{temporally} coarse-grained velocity of a particle.
We emphasize that the situation is different 
  from that of low-density fluids,
  in which the hydrodynamic velocity field is introduced 
  by \emph{spatial} coarse-graining of momentum 
  as a conserved quantity.
In deriving the Navier--Stokes equation for Newtonian fluids, 
  any confusion 
  between the velocity of a fluid element and the molecular velocity 
  should be severely criticized.
In dense liquids, however,
  the cage effect makes it possible---probably 
  as a crude approximation---%
  to regard the velocity of a particle  
  as the velocity 
  of a ``fluid element'' containing the particle, 
  with the understanding 
  that fluctuations at the collisional frequency
  should be filtered out mostly by time-averaging.

A simple way to remove the fast fluctuations in the velocity 
  is to focus on the displacement of the particle
  \cite{Einstein.AdPh17}.
Let us denote the position vector of the $i$-th particle 
  with $\mb{r}_i = \mb{r}_i(t)$;
  and its displacement, for the time interval from $s$ to $t$,
  with $\mb{R}_i = \mb{R}_i(t,s)$.
The displacement of the $i$-th particle
  is then given by integrating the velocity $\dot{\mb{r}}_i$
  over the time interval, 
  as 
\begin{equation}
  \mb{R}_i(t,s) = \mb{r}_i(t) - \mb{r}_i(s) 
  = \int_s^t {\dot{\mb{r}}_i(t')}\,\D{t'}
  \label{R=}.
\end{equation}

The most fundamental one-particle quantity 
  based on $\mb{R}_i(t,s)$ 
  is the mean square displacement (MSD), 
  which we write symbolically as $\Av{\mb{R}^2}$ 
  \cite{note.Av}.
On the assumption that the system is statistically steady,
  the MSD for ${t-s}\gg\tauA$
  grows asymptotically in proportion to $t-s$. 
This asymptotic behavior is understood 
  in terms of ``steps'' at $t_m = s + m\tauA$ ($m=0,1,\ldots$), 
  defined by $\mb{R}_i(t_{m+1}, t_m)$, 
  whose accumulation gives
\begin{equation}  
  \mb{R}_i(t_M,t_0) = \sum_{m=0}^{M-1} \mb{R}_i(t_{m+1}, t_m)
  \label{R//sum.step}; 
\end{equation}
  if separate steps are uncorrelated, 
  the MSD for the time interval from $t_0$ to $t_M$  
  (with $M$ being a positive integer)
  is proportional to $M = (t_M-t_0)/\tauA$.

Statistical quantities 
  based on displacements of two or more particles 
  are indicative of collective motions
  related to various aspects of the caged dynamics 
  \cite{Berthier.RMP83,Glotzer.JCP112,Ooshida.PRE88}.
In the presence of such collective motions,  
  in which the displacements of distinct particles 
  are spatially correlated,
  it seems justifiable 
  to define a \emph{displacement field} in some way.
Thus the hydrodynamical description 
  is extended to shorter scales
  without requiring momentum conservation. 
The existence of the displacement field 
  allows us to study the ``solidity'' of the liquid 
  by comparing the behavior of its displacement field 
  with that of visco\-elastic continuum models. 

\subsection{Displacement correlations}
\label{subsec:DC}

From among various statistical quantities 
  based on displacements in liquids,
  here we focus on one of its simplest form 
  involving two particles, 
  referring to it 
  as the (two-particle) 
  \emph{displacement correlation} (DC) tensor.
In terms of the particle displacement, 
  given in Eq.~(\ref{R=}), 
  the DC tensor is defined as follows
  \cite{Ooshida.BRL11,Ooshida.PRE94}: 
Denoting the Cartesian components of the displacement vector
  with 
\begin{equation}
  \mb{R}_i = \TwoVect{R_{ix}}{R_{iy}}  
  = \TwoVect{R_{ix}(t,s)}{R_{iy}(t,s)}\relax
\end{equation}
  in the 2D setup (and omitting the time arguments when obvious),
  we consider the tensorial product 
  of displacements of two particles ($i$ and $j$), 
\begin{equation}
  \mb{R}_i \otimes \mb{R}_j 
  =
  \begin{bmatrix}
    {R_{ix}}{R_{jx}} & {R_{ix}}{R_{jy}} \\
    {R_{iy}}{R_{jx}} & {R_{iy}}{R_{jy}} 
  \end{bmatrix}
  \relax.
\end{equation}
The DC tensor is then defined 
  by averaging $\mb{R}_i\otimes \mb{R}_j$
  over all pairs $(i,j)$ 
  such that 
  their relative position vector, at the ``initial'' time $s$, 
  equals a given vector $\di$. 
On the assumption 
  that the system is statistically homogeneous,
  hereafter 
  we denote the DC tensor with a capital $\chi$ as $\ChiR(\di,t,s)$, 
  making it clear 
  that the independent variables are $\di$, $t$ and $s$;
  then we write its definition
  symbolically as
\begin{equation}
  \ChiR(\di,t,s)
  = \Av{\mb{R}\otimes\mb{R}}_{\di}
  =
  \begin{bmatrix}
    \Av{{R_x}{R_x}}_{\di} & \Av{{R_x}{R_y}}_{\di} \\
    \Av{{R_y}{R_x}}_{\di} & \Av{{R_y}{R_y}}_{\di} 
  \end{bmatrix}
  \label{ChiR}, 
\end{equation}
  where $\Av{\quad}_{\di}$ denotes 
  conditional average over the pairs $(i,j)$ 
  satisfying $\mb{r}_j(s) - \mb{r}_i(s) = \di$; 
  see Eq.~(\ref{avr2p.observable}) in Appendix~\ref{app:DC}. 
For further details,
  see also Subsec.~III-A of Ref.~\cite{Ooshida.PRE94}.


On the assumption of statistical steadiness,
  the DCs depend on the time interval only through $\tD = t-s$.
Besides, 
  isotropy and reflectional symmetry are also assumed, 
  so that the DC tensor must be decomposable 
  into the longitudinal and the transverse correlations, 
  denoted with $\Xl$ and $\Xtr$, respectively.
We write this decomposition as 
\begin{multline}
  \ChiR(\di,t,s) = \ChiR(\tilde{\mb{d}},\tD) \brX  
  = \Xl (\di[\relax],\tD)  \frac{\di\otimes\di}{\di^2}
  + \Xtr(\di[\relax],\tD)
  \left( \openone - \frac{\di\otimes\di}{\di^2} \right)
  \label{ChiR.L+T},\hFil 
\end{multline}
  where $\di[\relax] = \abs{\smash{\di}}$.
Numerical procedure for calculating these correlations 
  from particle simulation data 
  is explained in Appendix~\ref{app:DC}.

\begin{figure}  
  \centering 
  \raisebox{5.7cm}{(a)}%
  \includegraphics[clip,height=6.0cm]{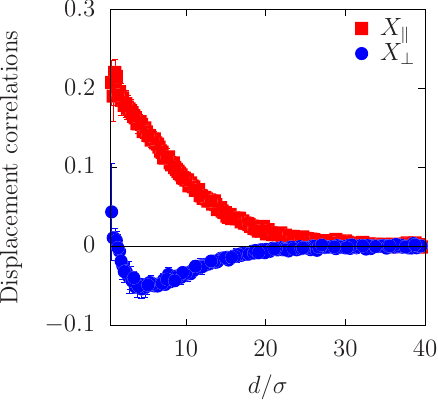}%
  \quad
  \raisebox{5.7cm}{(b)}\!%
  \includegraphics[clip,height=6.0cm]{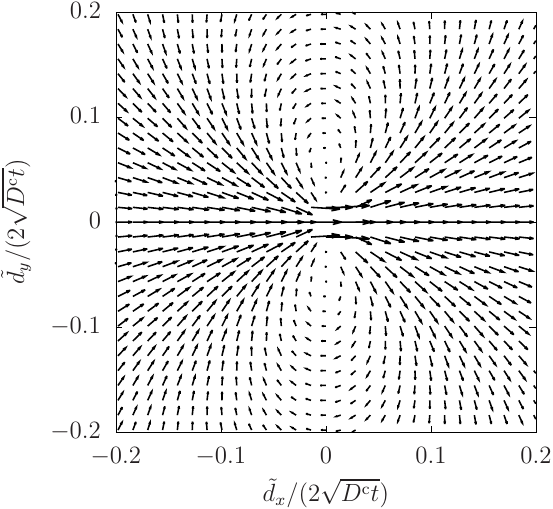}
  \caption{\label{Fig:RR.2D}%
    Typical behavior of DCs.
    The numerical conditions 
    are $\phiA = 0.5$ and $\tD = 20\tau_0$
    (the same as in Fig.~\protect\ref{Fig:cmp3} shown later); 
    see the main text for details.
    (a) 
    Longitudinal and transverse DCs, 
    plotted against initial distance $\di[\relax]$. 
    The particle diameter $\sigma$ is taken as a reference length scale.
    (b) 
    The numerical value of $\ChiR(\di,\tD) \cdot \mb{e}_1$, 
    plotted as a vector field on the $\di$-plane
    (in the fashion of Fig.~1 in Ref.~\protect\cite{Ooshida.PRE94},
    with $\mathbf{e}_1$
          denoting the ${x}$-directional unit vector).
    The axes are normalized with the diffusive length $2\sqrt{\Dc{t}}$
    appearing in Eq.~(\ref{sim.theta}).
  }
\end{figure}

Typical behavior of the DCs 
  as functions of $\di$, with $\tD$ fixed, 
  is depicted in Fig.~\ref{Fig:RR.2D}.
Details of the particle model used here \cite{Ooshida.PRE94}
  will be specified later in Sec.~\ref{sec:system}.
It is seen in Fig.~\ref{Fig:RR.2D}(a)
  that the two components of DC
  behave quite differently:
  $\Xl$ is positive everywhere, 
  while $\Xtr$ becomes negative for large $\di[\relax]$. %
This difference was noted 
  in a pioneering work by Doliwa and Heuer \cite{Doliwa.PRE61}, 
  many years before the researchers started to notice 
  that the transverse DC
  may contain information about shear modulus 
  of glassy systems \cite{Ikeda.PRE92,Flenner.PRL114}.
Essentially the same pattern of displacement correlations 
  is reported, seemingly independently, 
  as an experimental result by Cui \textit{et al.}~\cite{Cui.PRL92}, 
  who pointed out the presence of ``anti\-drag'' region 
  (where $\Xtr < 0$ in our notation).
The directional behavior of DCs 
  can be illustrated pictorially 
  as a pair of vortices \cite{Ooshida.PRE94,Doliwa.PRE61}, 
  as is exemplified in Fig.~\ref{Fig:RR.2D}(b).


Turning our attention 
  to the $\tD$-dependence of DCs,
  we find a rather curious fact:
  in comparison to $\tauA$
  determined as the timescale of $\Av{e^{\IK\cdot\mb{R}}}$,
  DCs are quite long-lived.
In the case of the ``vortices'' 
  reported by Doliwa and Heuer \cite{Doliwa.PRE61},  
  in a 2D system of Brownian particles (disks) 
  with the area fraction $\phiA = 0.77$, 
  the time interval chosen for their Fig.~8  
  was as long as $10\tauA$.
The same behavior of $\Xl$ and $\Xtr$ is observed
  even for colloidal liquids 
  which are only slightly glassy. 
In the case of $\phiA=0.5$ studied in Ref.~\cite{Ooshida.PRE94}, 
  the area fraction is so small that $\tauA$ is no longer 
  than the microscopic time scale $\tau_0 = \sigma^2/D$ 
  (composed of the disk diameter $\sigma$ 
  and the bare diffusivity $D$; see Sec.~\ref{sec:system}), 
  and yet the displacements are correlated 
  for much longer timescales,  
  as was shown 
  in Fig.~1 of Ref.~\cite{Ooshida.PRE94} ($\tD = 0.8\,\tau_0$), 
  which is now corroborated 
  in Fig.~\ref{Fig:RR.2D} of the present article ($\tD = 20\tau_0$). 

To explain the non-vanishing DC for $\tD\gg\tauA$,    
  Doliwa and Heuer \cite{Doliwa.PRE61}
  argued that 
  inter\-particle correlations from shorter times 
  can still contribute to DC for $\tD$.
If the two-particle displacements are correlated 
  only within $\tauA$,
  we can show 
  that the contributions to DC are accumulated 
  in proportion to $\tD$, 
  by following the same line of argument 
  as that for MSD with Eq.~(\ref{R//sum.step}).
We will demonstrate in Sec.~\ref{sec:num}, however, 
  that the actual $\tD$-dependence of DC 
  differs from the prediction of this simple scenario.
The numerical result suggests 
  that some aspects of the cage structure, probed by DC,  
  are more long-lived than $\tauA$.

\subsection{Shear strain correlations}
\label{subsec:SC}

The directionality and longevity of the DCs
  are reminiscent of the behavior of meso\-scopic shear strains
  \cite{Chikkadi.PRL107,Chattoraj.PRL111,Illing.PRL117}, 
  whose correlations has been recognized 
  as the Eshelby strain pattern 
  \cite{Eshelby.RSPA241,Picard.EPJE15} 
  depicted as a four-petaled flower-like figure.
We refer to this kind correlations  
  as \emph{shear strain correlations} 
  or, more concisely, \emph{shear correlation} (SC).

The four-petaled SC pattern 
  is often interpreted as a sign of elasticity, 
  showing the response of an elastic media 
  to localized plastic events.
Although the observations of the SC pattern 
  in sheared glasses 
  \cite{Chikkadi.PRL107,Chattoraj.PRL111}
  seems to be consistent with this interpretation, 
  it is curious that non-vanishing SC is reported experimentally
  also in quiescent glassy liquids, 
  being visible even for timescales longer than $\tauA$ 
  \cite{Illing.PRL117}.

Before studying the behavior of SC for $\tD \gg \tauA$, 
  here we review 
  how to calculate the SC from the particle simulation data.
A crucial assumption for definition of the SC 
  is the existence of the displacement field.  
Considering that there can be some choice 
  of independent variables in continuum mechanics, 
  here we choose $\mb{r}(s)$, 
  the position of the ``fluid element'' at the time $s$, 
  as the spatial independent variable.
Following basically the procedure   
  given by Illing \textit{et al.}~\cite{Illing.PRL117}
  in their supplementary material, 
  we define 
  the displacement field 
  $\mb{R} = \mb{R}(\mb{r}(s),t,s)$ 
  as 
\begin{equation}
  \mb{R}(\mb{r}(s),t,s)
  = \Ev{
  \frac{1}{\rho(\mb{x},s)}
  \sum_i \mb{R}_i(t,s) \de(\mb{r}_i(s) -\mb{x})}{\mb{x}=\mb{r}(s)}
  \relax,
\end{equation}   
  where $\de(~\cdot~)$ is a slightly blurred delta function \cite{note.deltaE}, 
  and $\rho$ is the density field defined as 
\begin{equation}
  \rho(\mb{x},t) = \sum_i \de(\mb{r}_i(t) -\mb{x})
  \label{rho=}. 
\end{equation}
Taking the statistical steadiness into account,
  we can write $\mb{R}(\mb{r}(s),t,s) = \mb{R}(\mb{r}(s),\tD)$.
Once $\mb{R}$ is defined, 
  the relative deformation for the time interval from $s$ to $t$ 
  is given by the mapping 
\begin{equation}
  \mb{r}(s) = \TwoVect{x(s)}{y(s)} 
  \mapsto 
  \mb{r}(t) = \TwoVect{x(t)}{y(t)}
  = \mb{r}(s) + \mb{R}(\mb{r}(s),\tD) 
  \relax.  
\end{equation}
The deformation gradient
  relative to the configuration at the time $s$, 
  which we denote with $\mathsf{F}$, 
  is then given by 
\begin{equation}
  \mathsf{F} = \dd{\mb{r}(t)}{\mb{r}(s)}
  = 
  \begin{bmatrix}
  \rD{x(t)}/\rD{x(s)} & \rD{x(t)}/\rD{y(s)}  \\
  \rD{y(t)}/\rD{x(s)} & \rD{y(t)}/\rD{y(s)}  
  \end{bmatrix}
  \label{F=dr@t/dr@s}.
\end{equation}
The symmetric part of $\mathsf{F}$ 
  gives the strain tensor. 
In particular, 
  the sum of the off-diagonal components,
  which we denote with 
\begin{equation}
  \gm = \gm(\mb{r}(s),t,s)
  = F_{12} + F_{21} = \dd{x(t)}{y(s)} + \dd{y(t)}{x(s)}
  \label{gm=}, 
\end{equation}
  would correspond to the shear strain 
  if the system would be driven along the $x$-axis
  so that $\Av{\mb{R}(\mb{r}(s),\tD)} \propto (y(s)\tD,0)$.
In the absence of external driving, 
  still we refer to $\gm$ in Eq.~(\ref{gm=})
  as the shear strain field. 
The SC is defined as correlation of $\gm$ 
  at two positions separated by $\di$
  (say, at $\mb{x}'$ and $\mb{x}' + \di$) \cite{note.Av}: 
\begin{equation}
  \chi_\gm^{} = \Av{\gm\gm}_{\di}
  = \Av{\gm({\mb{x}'},t,s) \gm({\mb{x}'} +\smash{\di},t,s)} 
  \label{chiGm=}.
\end{equation}
Due to the spatial and temporal translational symmetry
  of the system, 
  $\chi_\gm^{}$ 
  depends on the space--time interval $(\di,\tD)$ alone, 
  and not on $\mb{x}'$ and $s$ directly.

\begin{figure}
\includegraphics[clip,width=7.0cm]{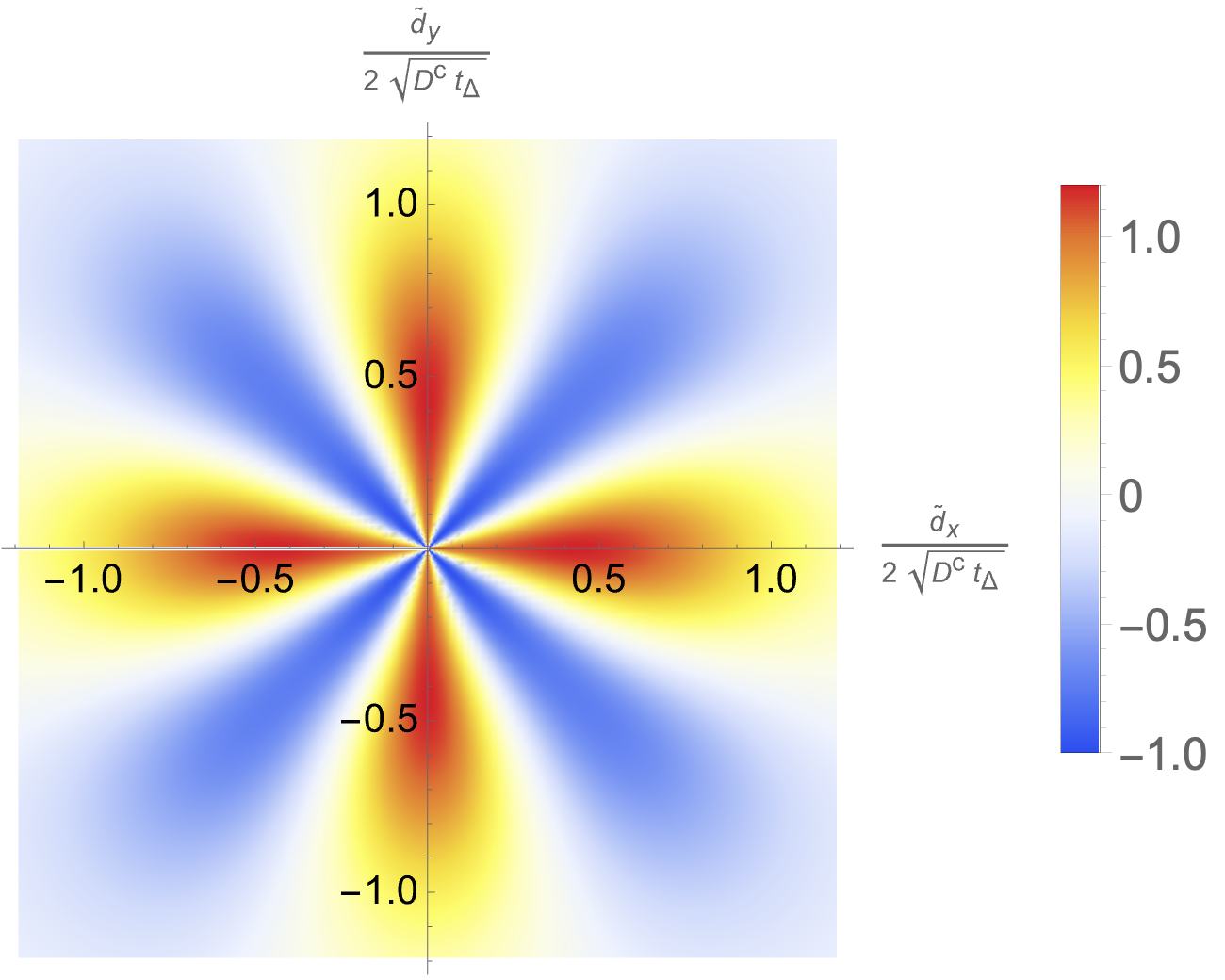}
\caption{\label{Fig:chiGm}%
  Shear correlation $\chi_\gm^{}$ 
  given by Eq.~(\protect\ref{chiGm//th})
  with $\muR = 0.25$. 
  On the understanding 
  that $\vartheta \propto \xi_* \propto \abs{\smash{\di}}$, 
  the value of $\chi_\gm^{}/(A\di^2)$ 
  is plotted as a color map on the $\di$-plane, 
  being normalized with the factor $A$ 
  to fit within the range from $-1$ to $+1$.
  The axes are normalized with the diffusive length $2\sqrt{\Dc{t}}$
  appearing in Eq.~(\ref{sim.theta}).
}
\end{figure}

In order to answer the questions about the correlations 
  raised in Introduction, 
  we will make comparative studies of DC and SC 
  in the following sections. 
In particular,
  a formula interconnecting the DC and the SC 
  will be developed in Sec.~\ref{sec:relation}. 
As a usage demonstration of the formula, 
  we can obtain an approximate analytical expression for SC,
  later given as Eq.~(\ref{chiGm//th}),  
  from that of DC given in Ref.~\cite{Ooshida.PRE94}. 
The result is shown in Fig.~\ref{Fig:chiGm}
  as a color map on the $\di$-plane, 
  in which the four-petal flower-like pattern is evident.

\section{Relations involving shear strain correlations}
\label{sec:relation}

As we reviewed in the previous section, 
  the DC and the SC 
  share a common ingredient, 
  in the sense 
  that both are calculated from the displacements of the particles.
The two kinds of correlations 
  have something in common, for example, 
  in that they reflect the vectorial character of the displacement, 
  and also in that they can persist longer than $\tauA$.

These observations have led us to ask
  whether the two correlations are equivalent to each other.
Although the displacement is used 
  as a common ingredient for the DC and the SC, 
  we have a reason to doubt their equivalence: 
  there is an evident characteristic length 
  in the transverse DC in Fig.~\ref{Fig:RR.2D}, 
  marked by the point at which $\Xtr$ changes the sign
  or by the minimum of $\Xtr$, 
  while such a length scale is not so apparent in SC. 

In order to answer this question, 
  let us develop a theoretical framework 
  to derive an analytical relation between the DC and the SC. 
Readers who are more interested in the relation itself 
  than its derivation procedure 
  may skim the theoretical development in Subsec.~\ref{subsec:psi//xi}
  and, after checking some basic notation 
  in and around 
  Eqs.~(\ref{RR*})--(\ref{polar}), (\ref{gm.0+4}) and (\ref{f//RR}), 
  jump to Eq.~(\ref{gg//RR.polar}).

\subsection{Label variable and deformation gradient tensor field}
\label{subsec:psi//xi}

In order to relate the DC with SC analytically, 
  we need to formulate the two kinds of correlations 
  on the same ground, 
  taking the analytical amenability into account.
For this aim, 
  we adopt the \emph{label variable} formulation 
  developed in our previous works on the DC tensor 
  \cite{Ooshida.BRL11,Ooshida.PRE94}.
  
The basic idea is 
  to introduce the label variable $\BoldXi = (\xi_1,\xi_2)$
  convected by the velocity field $\mb{u}$, 
  which means 
  that $\xi_\Ix = \xi_\Ix(\mb{x},t)$ ($\Ix\in\{1,2\}$) 
  satisfies 
\begin{equation}
  \left( \dt + \mb{u}\cdot\dd{}{\mb{x}} \right) 
  \xi_\Ix(\mb{x},t) = 0   
  \label{convect.xi}
\end{equation}  
  where 
\begin{equation}
  \mb{u} = \mb{u}(\mb{x},t) 
  = \frac{\mb{Q}(\mb{x},t)}{\rho(\mb{x},t)}
  \relax,\quad
  \mb{Q}(\mb{x},t) 
  = \sum_i \dot{\mb{r}}_i(t) \de(\mb{r}_i(t) -\mb{x})
  \relax, 
\end{equation}  
  and $\rho(\mb{x},t)$ 
  is the density field given by Eq.~(\ref{rho=}).
From among infinitely many solutions to Eq.~(\ref{convect.xi})
  corresponding to different initial data, 
  we choose a solution that satisfies 
  $\partial(\xi_1,\xi_2)/\partial(x_1,x_2) = \rho(\mb{x},t)$
  with the aid of the continuity equation 
  \cite{Ooshida.BRL11},
  thus constructing $\BoldXi = \BoldXi(\mb{x},t)$ 
  subject to Eq.~(\ref{convect.xi}) at each instant 
  and normalized so that 
\[
  \abs{\BoldXi(\mb{r}_i(t),t) - \BoldXi(\mb{r}_j(t),t)}
  \simeq \frac{\abs{\mb{r}_i(t) - \mb{r}_j(t)}}{\ell_0} 
  \relax;
\]
  here $\ell_0 = 1/\sqrt{\rho_0}$ 
  is the typical inter\-particle distance, 
  defined in terms of the mean density $\rho_0$.
Then, by taking $\BoldXi$ as the independent variable, 
  we transfer to a $t$-dependent curvilinear coordinate system 
  sticking to the particles
  (known by the name of the convective coordinate system), 
  in the form of a mapping
\begin{equation}
  \BoldXi \mapsto \mb{r} = \mb{r}(\BoldXi,t) = \TwoVect{x}{y}
  \label{xi-to-r}
\end{equation}
  such that $\mb{r}(\BoldXi(\mb{x},t),t) = \mb{x}$.

While the partial derivative of $\mb{r} = \mb{r}(\BoldXi,t)$ 
  with regard to $t$ 
  gives the velocity $\mb{u}$, 
  its partial derivative with regard to $\BoldXi$
  yields what is called 
  the deformation gradient tensor \cite{Marsden.Book1994}
  or the displacement gradient tensor \cite{Bird.Book1987}.
It is convenient 
  to rearrange the components of this tensor field, 
  $\rD\mb{r}/\rD\BoldXi$,   
  into a form 
  corresponding to the Helmholtz decomposition 
  of the displacement field \cite{Klix.PRL109,Ooshida.PRE94}, 
  which consists of the dilatational component $\PsiD$ 
  and the rotational component $\PsiR$.
We write these components as
\begin{subequations}%
\begin{align}
  \PsiD(\BoldXi,t) 
  &= \ell_0^{-1} \left( \dXi{x} + \dHa{y} \right) - 2 
  \label{psiD}, \\
  \PsiR(\BoldXi,t) 
  &= \ell_0^{-1} \left( \dXi{y} - \dHa{x} \right) 
  \label{psiR}, 
\end{align}%
\label{eqs:Psi}%
\end{subequations}%
  using abbreviation
  $\partial_\Ix = \partial/\partial{\xi_\Ix}$
  ($\Ix \in \{1,2\}$).

Expanding $\PsiD(\BoldXi,t)$ and $\PsiR(\BoldXi,t)$  
  into Fourier series as 
\begin{equation}
  \Psi_a(\BoldXi,t) 
  = \sum_{\mb{k}} \psi_a(\K,t) e^{-\IK\cdot\BoldXi} 
  \quad (a\in\{\mathrm{d},\mathrm{r}\})
  \label{psi.Fourier}, 
\end{equation}  
  we define correlations of the Fourier modes
  by 
\begin{equation}
  C_a(\K,t,s)  %
  = \mathcal{N}\Av{{\psi_a(\K,t)}{\psi_a(-\K,s)}}  
  \label{CorrH=},
\end{equation}
  where $\mathcal{N}$ is a normalizing constant;
  here we choose $\mathcal{N}$  
  equal to the total particle number $N$ 
  in the domain.

We expect that the correlations 
  of the deformation gradient field, 
  namely $\Cd$ and $\Cr$, 
  carry information about stresses 
  associated with their respective modes of deformation.
These stresses 
  can be related to the elastic moduli
  as was discussed by Klix \textit{et al.}~\cite{Klix.PRL109},
  and they can be viscoelastic or simply viscous.
The problem is 
  how this information of rheology 
  is reflected by the behavior of DC and SC.

The displacement of the fluid element labeled by $\BoldXi$
  is given by 
\begin{equation}
  \mb{R}(\BoldXi,t,s) = \mb{r}(\BoldXi,t) - \mb{r}(\BoldXi,s)
  \label{R//xi}
\end{equation}
  on the basis of the mapping in Eq.~(\ref{xi-to-r}).
Using Eq.~(\ref{R//xi}), 
  we can derive a formula
  for calculating the DC tensor from $\Cd$ and $\Cr$ %
  \cite{Alexander.PRB18,Ooshida.PRE94}. 
The formula will be given 
  later as Eqs.~(\ref{eqs:I.AP2}).

In relating the SC with $\Cd$ and $\Cr$, 
  some caution is needed 
  to avoid confusion between $\gm$ and $\dHa{x} + \dXi{y}$; 
  the former, $\gm = F_{12} + F_{21}$, 
  is originally a two-time quantity,  
  as it comes from $\mathsf{F}$ in Eq.~(\ref{F=dr@t/dr@s})
  which depends both on $s$ and on $t$, 
  while the latter is a one-time quantity 
  coming from $\partial\mb{r}(\BoldXi,t)/\partial\BoldXi$.
More suitably, 
  we notice 
  that $\mathsf{F}$ is the differential quotient 
  of the composite mapping 
\begin{equation}
  \mb{r}(s) 
  \mapsto \BoldXi = \BoldXi(\mb{r}(s), s)
  \mapsto \mb{r}(t) = \mb{r}(\BoldXi,t) 
\end{equation}
  in which $\mb{r}(t)$ and $\mb{r}(s)$ are connected
  by way of $\BoldXi$ as a parameter.
Denoting the components of the relative deformation gradient tensor 
  with $F_{\alpha\beta} = \rD{r_\alpha}(t)/\rD{r_\beta}(s)$, 
  we have $\D{r_\alpha}(t) = F_{\alpha\beta}\D{r_\beta}(s)$,
  so that Eq.~(\ref{F=dr@t/dr@s}) can be written, 
  with $\BoldXi$ taken as the independent variable, 
  as
\begin{subequations}%
  \begin{equation}
  \dd{\mb{r}(\BoldXi,t)}{\BoldXi}
  = \mathsf{F}\cdot\dd{\mb{r}(\BoldXi,s)}{\BoldXi}
\end{equation}
  or, in terms of components,
\begin{equation}
  \begin{bmatrix}
  \dXi{x} & \dHa{x}\\
  \dXi{y} & \dHa{y} 
  \end{bmatrix}_t
  = 
  \begin{bmatrix} F_{11} & F_{12} \cr F_{21} & F_{22} \end{bmatrix}
  \begin{bmatrix}
  \dXi{x} & \dHa{x}\\
  \dXi{y} & \dHa{y} 
  \end{bmatrix}_s
  \relax, 
\end{equation}%
\label{eqs:PsiT=F*PsiS}%
\end{subequations}%
  with subscripted $t$ or $s$ denoting the time argument.
Note that $\rD\mb{r}/\rD\BoldXi$ in Eq.~(\ref{eqs:PsiT=F*PsiS}) 
  is connected to $\PsiD$ and $\PsiR$ through Eq.~(\ref{eqs:Psi}).
With this connection taken into account,  
  Eq.~(\ref{eqs:PsiT=F*PsiS}) 
  provides a foundation for calculation of the SC, 
  as $\gm = F_{12} + F_{21}$.

\subsection{Formulae between the correlations under consideration}
\label{subsec:formula}

On the basis of the label variable formulation 
  of the displacement field $\mb{R}(\BoldXi,t,s)$
  in the previous subsection,
  here we interrelate 
  three kinds of correlations: 
  the DC, the SC,
  and the correlations of the deformation gradients 
  (namely $\Cd$ and $\Cr$). 
We start by reviewing 
  a formula connecting $\Cd$ and $\Cr$ to the DC 
  \cite{Ooshida.BRL11,Ooshida.PRE94}.
Subsequently, 
  we perform analogous calculations for the SC, 
  reproducing the formula 
  reported by Illing \textit{et al.}~\cite{Illing.PRL117}. 
Using relations 
  obtained in course of these calculations, 
  we derive a relation interconnecting the DC and the SC.

\subsubsection{Deformation gradient correlations to DC}

Let us begin with a formula 
  to calculate the DC tensor, 
  $\ChiR = \Av{\mb{R}\otimes\mb{R}}_{\di}$, 
  from the deformation gradient correlations, 
  $\Cd$ and $\Cr$.
The starting point 
  is the displacement field $\mb{R}(\BoldXi,t,s)$ 
  in Eq.~(\ref{R//xi}).
On the understanding 
  that the initial separation $\di$ in the physical space
  is equivalent to the label-space separation $\XiR = \di/\ell_0$
  on average, 
  we calculate the DC as 
\begin{align}
  \ChiR(\di,t,s) 
  &\simeq 
  \Av{\mb{R}(\BoldXi,t,s)\otimes \mb{R}(\BoldXi',t,s)
  }_{\BoldXi-\BoldXi'=\di/\ell_0}
  \notag \\
  &= \Av{\mb{R}(\BoldXi'+\XiR,t,s) \otimes \mb{R}(\BoldXi',t,s)}
  \label{RR*}.
\end{align}   
Note that the expression on the right side of Eq.~(\ref{RR*}) 
  is actually independent of $\BoldXi'$ 
  due to the space-translational symmetry.

By expressing the displacement field $\mb{R}(\BoldXi,t,s)$
  in terms of the Fourier modes 
  of the deformation gradient tensor, 
  we can derive a formula 
  that relates the DC tensor 
  to the correlations of $\psi_a$,  
  where $a\in\{\mathrm{d},\mathrm{r}\}$
  \cite{Ooshida.BRL11,Ooshida.PRE94}.
We refer to it as the Alexander--Pincus formula, 
  naming it after the authors of Ref.~\cite{Alexander.PRB18}.
In writing this formula,
  it is convenient 
  to introduce $C_a^0(\K,s) = C_a(\K,s,s)$
  and 
\begin{align}
  C^\Delta_a(\K,t,s) 
  &= 
  \frac{C_a^0(\K,s) + C_a^0(\K,t)}{2} - C_a(\K,t,s)  
  \notag \\  
  &= 
  {\frac{N}{2}} \Av{%
  \abs{{\psi_a( \K,t)} - {\psi_a( \K,s)}}^2
  }
  \label{C.Delta}.
\end{align}
The Alexander--Pincus formula 
  then reads \cite{Ooshida.BRL11,Ooshida.PRE94}
\begin{subequations}
\begin{equation}
  \ChiR(\ell_0\XiR,t,s)  
  = \frac{\ell_0^2}{2\pi^2} \Id 
  + \frac{\ell_0^2}{2\pi^2} \Ir 
  \label{AP2.Id+Ir}, 
\end{equation}
  where 
\begin{align}
  \Id  &= 
  \iint \Cd^\Delta(\K,t,s)
  \begin{bmatrix}
  k_1^2          & {k_1}{k_2} \\
  {k_2}{k_1} & k_2^2
  \end{bmatrix}
  \frac{e^{-\IK\cdot\XiR}}{\K^4}  \D{k_1}\D{k_2}
  \relax,
  \\
  \Ir  &= 
  \iint \Cr^\Delta(\K,t,s)
  \begin{bmatrix}
  k_2^2           & \!{-{k_1}{k_2}} \\
  -{k_2}{k_1} &          k_1^2
  \end{bmatrix}
  \frac{e^{-\IK\cdot\XiR}}{\K^4}  \D{k_1}\D{k_2}
  \relax.
\end{align}%
\label{eqs:I.AP2}%
\end{subequations}

\begin{widetext}%
The assumption of statistical isotropy 
  implies $C_a^\Delta(\K,t,s) = C_a^\Delta(k,t,s)$
  and suggests to introduce polar coordinates 
  in the $\K$-space and the $\XiR$-space, 
\begin{equation}
  \K = \TwoVect{k_1}{k_2} 
  = k \TwoVect{\cos\varphi}{\sin\varphi}
  \relax, \quad 
  \XiR = \TwoVect{\xi_{*1}}{\xi_{*2}} 
  = \xi_* \TwoVect{\cos\varphi_*}{\sin\varphi_*}
  \label{polar}, 
\end{equation}
  so that we can rewrite the integrals in Eqs.~(\ref{eqs:I.AP2})
  as 
\begin{subequations}
\begin{align}
  \Id  
  &= 
  \iint \Cd^\Delta(k,t,s)
  \begin{bmatrix}
  \cos^2\varphi            & \cos\varphi\sin\varphi \\
  \,\sin\varphi\cos\varphi & \sin^2\varphi 
  \end{bmatrix}
  \frac{e^{-\Ik\xi_* \cos(\varphi - \varphi_*)}}{k}\,
  \D{k}\,\D{\varphi}
  \notag \\ 
  &= 
  \pi \int_0^\infty 
  \Cd^\Delta(k,t,s) \left\{
  \begin{bmatrix} 1 & 0 \\ 0 & 1 \end{bmatrix}
  J_0(k\xi_*) 
  -  
  \begin{bmatrix}
  \cos{2\varphi_*}  & \,\sin{2\varphi_*} \\
  \sin{2\varphi_*}  & - \cos{2\varphi_*} 
  \end{bmatrix}
  J_2(k\xi_*)
  \right\}
  \frac{\D{k}}{k}
  \label{Id.polar},
  \\
  \Ir  
  &= 
  \iint \Cr^\Delta(k,t,s)
  \begin{bmatrix}
  \sin^2\varphi           &\!{-\cos\varphi\sin\varphi} \\
  -\sin\varphi\cos\varphi &        \cos^2\varphi 
  \end{bmatrix}
  \frac{e^{-\Ik\xi_* \cos(\varphi - \varphi_*)}}{k}\,
  \D{k}\,\D{\varphi}
  \notag \\ 
  &= 
  \pi \int_0^\infty 
  \Cr^\Delta(k,t,s) \left\{
  \begin{bmatrix} 1 & 0 \\ 0 & 1 \end{bmatrix}
  J_0(k\xi_*) 
  + 
  \begin{bmatrix}
  \cos{2\varphi_*}  & \,\sin{2\varphi_*} \\
  \sin{2\varphi_*}  & - \cos{2\varphi_*} 
  \end{bmatrix}
  J_2(k\xi_*)
  \right\}
  \frac{\D{k}}{k}
  \label{Ir.polar}, 
\end{align}%
\label{eqs:AP2.polar}%
\end{subequations}%
  with the Bessel functions, $J_0$ and $J_2$, 
  arising from integration over $\varphi$ \cite{Arfken.Book2013}. 
This form can be more useful than Eqs.~(\ref{eqs:I.AP2})
  in evaluating the integrals 
  when $\Cd(k,t,s)$ and $\Cr(k,t,s)$ are known.  
\end{widetext}

\subsubsection{Deformation gradient correlations to SC}

In parallel to the Alexander--Pincus formula 
  that expresses the DC 
  in terms of $\Cd^\Delta$ and $\Cr^\Delta$, 
  it is possible to derive an analogous formula for the SC. 
The derivation 
  starts with solving Eq.~(\ref{eqs:PsiT=F*PsiS}) 
  for $\mathsf{F}$, 
  which gives 
\begin{equation}
  \mathsf{F} 
  = \begin{bmatrix} F_{11} & F_{12} \\ F_{21} & F_{22} \end{bmatrix}
  = 
  \begin{bmatrix}
  \dXi{x} & \dHa{x}\\
  \dXi{y} & \dHa{y} 
  \end{bmatrix}_t
  \left( \rho 
  \begin{bmatrix}
  ~\;\dHa{y} & \! {- \dHa{x}}\\
  - \dXi{y} & \! {~\;\dXi{x}}
  \end{bmatrix}
  \right)_s
\end{equation}
  with the Jacobian determinant 
  $\partial(x,y)/\partial(\xi_1,\xi_2) = 1/\rho$ 
  taken into account.
The time arguments are subscripted  
  as in Eqs.~(\ref{eqs:PsiT=F*PsiS}).
The off-diagonal components of $\mathsf{F}$
  are then evaluated:
\begin{subequations}%
\begin{align}
  F_{12} &= \rho_s 
  \left( -\Ev{\dd{x}{\xi_1}}{t} \Ev{\dd{x}{\xi_2}}{s}
  +       \Ev{\dd{x}{\xi_2}}{t} \Ev{\dd{x}{\xi_1}}{s} \,\right) 
  \br\simeq \ell_0^{-1} 
  \left( \dHa{x(\BoldXi,t)} - \dHa{x(\BoldXi,s)} \right)
  \relax, 
  \\
  F_{21} &= \rho_s 
  \left (~ \Ev{\dd{y}{\xi_1}}{t} \Ev{\dd{y}{\xi_2}}{s}
  -        \Ev{\dd{y}{\xi_2}}{t} \Ev{\dd{y}{\xi_1}}{s} \,\right) 
  \br\simeq \ell_0^{-1} 
  \left( \dXi{y(\BoldXi,t)} - \dXi{y(\BoldXi,s)} \right)
  \relax,
\end{align}%
\end{subequations}%
  where the lowest-order approximation 
  for the diagonal components,
  $\dXi{x} \simeq \dHa{y} \simeq \ell_0$,
  is taken into account \cite{note.1+Psi}.
Upon substitution into the definition of $\gm$,
  namely Eq.~(\ref{gm=}), 
  we find   
\begin{multline}
  \gm = F_{12} + F_{21} \brX 
  \simeq \ell_0^{-1} \left( 
  {}\dHa{x(\BoldXi,t)} - \dHa{x(\BoldXi,s)} 
  + \dXi{y(\BoldXi,t)} - \dXi{y(\BoldXi,s)} 
  \right)
  \label{gm//Psi},\hFil  
\end{multline}
  which is simplified 
  by using the displacement field, 
  $\mb{R} = \mb{r}(\BoldXi,t) - \mb{r}(\BoldXi,s)$, 
  as 
\begin{equation}
  \gamma = \ell_0^{-1} \left( \dXi{R_y} + \dHa{R_x} \right)
  \label{gm//R}.
\end{equation}

The expression of $\gm$ in Eq.~(\ref{gm//R})
  makes it possible to calculate the SC 
  by evaluating 
\begin{equation}
  \chi_\gm^{} = \chi_\gm^{}(\ell_0\XiR,t,s) 
  = \Av{\gm(\BoldXi'+\XiR,t,s)\gm(\BoldXi',s,t)}
  \label{gm@xi*gm@0}, 
\end{equation}
  which is supposed to result in an integral form 
  analogous to Eq.~(\ref{eqs:I.AP2}) for DC.
In fact, 
  it reproduces a formula 
  briefly stated by Illing \textit{et al.}~\cite{Illing.PRL117},
  which reads
\begin{align}
  \chi_\gm^{}
  &=
  \frac{1}{{2\pi^2}}
  \iint 
  \Cd^\Delta(\K,t,s) 
  \frac{4{k_1^2}{k_2^2}}{\K^4}   e^{-\IK\cdot\XiR} \D{k_1}\D{k_2}
  \br 
  + \frac{1}{{2\pi^2}}
  \iint 
  \Cr^\Delta(\K,t,s) 
  \frac{(k_1^2 - k_2^2)^2}{\K^4} e^{-\IK\cdot\XiR} \D{k_1}\D{k_2}
  \label{gm//C.H}
\end{align}
  in our notation.

To evaluate Eq.~(\ref{gm@xi*gm@0}), 
  we operate Eq.~(\ref{gm//R}) with ${\dXi}^2 + {\dHa}^2$ 
  and rearrange the terms
  so as to express the result 
  in terms of $\PsiD$ and $\PsiR$.
Taking notice of the relations
\begin{subequations}
\begin{align}
  &\dXi{R_x} + \dHa{R_y} = \PsiD(\BoldXi,t) - \PsiD(\BoldXi,s), \\
  &\dXi{R_y} - \dHa{R_x} = \PsiR(\BoldXi,t) - \PsiR(\BoldXi,s),
\end{align}%
\end{subequations}%
  we find
\begin{widetext}%
\begin{align}
  \left( {\dXi}^2 + {\dHa}^2 \right) \gm 
  &= \ell_0^{-1} \left[
  2 \dXi\dHa          (\dXi{R_x} + \dHa{R_y})
  + (\dXi^2 - \dHa^2) (\dXi{R_y} - \dHa{R_x}) 
  \right] \notag \\
  &= \ell_0^{-1} \left\{
  2 \dXi\dHa{}        [ \PsiD(\BoldXi,t) - \PsiD(\BoldXi,s) ]
  + (\dXi^2 - \dHa^2) [ \PsiR(\BoldXi,t) - \PsiR(\BoldXi,s) ]
  \right\}
  \relax, 
\end{align}
  which is rewritten in Fourier representation 
  as 
\begin{equation}
  \gamma(\BoldXi,t,s) 
  = \sum_{\K}
  \frac{e^{-\IK\cdot\BoldXi}}{\K^2}
  \left\{
  2 {k_1}{k_2} 
  \left[ \psiD(\K,t) - \psiD(\K,s) \right] 
  +
  \left( k_1^2 - k_2^2 \right)  
  \left[ \psiR(\K,t) - \psiR(\K,s) \right] 
\right\}
\label{gm//psiH}.
\end{equation}
\end{widetext}%
Substituting Eq.~(\ref{gm//psiH}) into Eq.~(\ref{gm@xi*gm@0}), 
  assuming that the correlations 
  between different modes are negligible
  (by the same reason as is discussed in derivation 
  of the Alexander--Pincus formula \cite{Ooshida.BRL11}), 
  and taking the continuum limit, 
  we arrive at Eq.~(\ref{gm//C.H}).

The four-fold symmetry of $\chi_\gamma$ 
  is readily shown 
  by rewriting Eq.~(\ref{gm//C.H}) in polar coordinates.
Using the angular variables $\varphi$ and $\varphi_*$ 
  in Eq.~(\ref{polar}), 
  we have 
\begin{gather*}
  \frac{4{k_1^2}{k_2^2}}{\K^4}   
  = \sin^2{2\varphi} = \frac{1-\cos{4\varphi}}{2} 
  \relax, \\
  \frac{(k_1^2 - k_2^2)^2}{\K^4} 
  = \cos^2{2\varphi} = \frac{1+\cos{4\varphi}}{2}
  \relax, \\
  e^{-\IK\cdot\XiR}
  = e^{-\Ik\xi_*\cos(\varphi-\varphi_*)} 
\end{gather*}
  for the $\varphi$-dependent factors in the integrands.
Subsequently, 
  changing the variable of integration 
  from $\varphi$ to $\varphi-\varphi_*$, 
  we find the $\varphi_*$-dependence of $\chi_\gm^{}$ 
  to be of the form
\begin{equation}
  \chi_\gm 
  = \chi_\gm^{(0)} + \chi_\gm^{(4)} \cos{4\varphi_*}
  \label{gm.0+4}
\end{equation}
  where
\begin{subequations}
\begin{align}
  \chi_\gm^{(0)} 
  &= {\frac{1}{2\pi}}\int_0^\infty 
  \left[  \Cd^\Delta(\K,t,s) + \Cr^\Delta(\K,t,s) \right]
  J_0(k\xi_*) k\D{k}\relax,
  \\
  \chi_\gm^{(4)} 
  &= {\frac{1}{2\pi}}\int_0^\infty 
  \left[ -\Cd^\Delta(\K,t,s) + \Cr^\Delta(\K,t,s) \right]
  J_4(k\xi_*) k\D{k}
\end{align}%
\label{eqs:gm04//int}%
\end{subequations}%
  are independent of $\varphi_*$.
This is almost purely 
  a consequence of the definitions of SC, 
  independent of the behavior of $\Cd^\Delta$ and $\Cr^\Delta$.
To be more precise,
  we have assumed 
  only the existence of the label-based displacement field  
  and the isotropy of the system.

\subsubsection{Relation between DC and SC}

In addition to Eq.~(\ref{eqs:I.AP2}) for DC 
  and Eq.~(\ref{gm//C.H}) for SC, 
  now we derive a third relation
  expressing the SC in terms of $\Xl$ and $\Xtr$.
Such a relation must be available
  in principle, 
  because the Alexander--Pincus formula is invertible 
  in the sense that $\Cd^\Delta$ and $\Cr^\Delta$ 
  can be expressed in terms of the DC \cite{Ooshida.PRE94}, 
  and substitution into Eq.~(\ref{gm//C.H})
  then yields $\chi_\gm^{}$. 

To derive the relation more straightforwardly, 
  we begin with substituting Eq.~(\ref{gm//R}), 
  expressing $\gm$ in terms of $\mb{R}(\BoldXi,t,s)$, 
  directly into Eq.~(\ref{gm@xi*gm@0}) that gives $\chi_\gm^{}$.
Subsequently, 
  taking the space-translational symmetry into account,
  we rewrite the expression for SC 
  as 
\begin{widetext}%
\begin{align}
  \chi_\gm^{} 
  &= \ell_0^{-2}\Av{\vphantom{\sqrt{\mathstrut}}%
  [ \dXiPrime{R_y}(\BoldXi'+\XiR) + \dHaPrime{R_x}(\BoldXi'+\XiR) ]
  [ \dXiPrime{R_y}(\BoldXi')      + \dHaPrime{R_x}(\BoldXi')      ]  
  }
  \notag \\   
  &= -\ell_0^{-2}
  \Av{\vphantom{\sqrt{\mathstrut}}\left\{%
  \dXiPrime{
  [ \dXiPrime{R_y}(\BoldXi'+\XiR) + \dHaPrime{R_x}(\BoldXi'+\XiR) ]
  }\right\}{R_y}(\BoldXi') }   
  -\ell_0^{-2}
  \Av{\vphantom{\sqrt{\mathstrut}}\left\{%
  \dHaPrime{
  [ \dXiPrime{R_y}(\BoldXi'+\XiR) + \dHaPrime{R_x}(\BoldXi'+\XiR) ]
  }\right\}{R_x}(\BoldXi') } 
  \label{gm@x1*gm@x2}
\end{align}%
\end{widetext}%
  where $\partial_{\Ix'}$ 
  stands for $\partial/\partial{\xi'_\Ix}$ ($\Ix \in \{1,2\}$). 
In Eq.~(\ref{gm@x1*gm@x2}),
  use is made of ``integration by parts''
\[  
  \Av{\psi\partial_{\Ix'}\phi} = -\Av{(\partial_{\Ix'}\psi)\phi}
\]  
  for arbitrary functions 
  $\psi = \psi(\BoldXi')$ and $\phi = \phi(\BoldXi')$ 
  such that $\Av{\psi\phi}$ is spatially uniform.

Manipulating the differentiations in Eq.~(\ref{gm@x1*gm@x2}) 
  with relations such as $\dXiPrime{\mb{R}(\BoldXi'+\XiR)} 
  = \dXiAst {\mb{R}(\BoldXi'+\XiR)}$ 
  (with the meaning of $\partial_{\Ix_*}$ obviously understood),
  and using the ``double-dot'' product notation 
  \cite{Morse.Book1953},
  we obtain 
\begin{equation}
  \chi_\gm^{}  
  = -\ell_0^{-2}
  \begin{bmatrix}
  \dXiAst \dXiAst  & \dXiAst \dHaAst       \\
  \dHaAst \dXiAst  & \dHaAst \dHaAst   
  \end{bmatrix}
  :
  \begin{bmatrix}
  \Av{{R_y}{R_y}} & \Av{{R_x}{R_y}} \\
  \Av{{R_y}{R_x}} & \Av{{R_x}{R_x}}  
  \end{bmatrix}  
  \label{gg//RR}
\end{equation}  
  where 
  $\Av{{R_x}{R_x}} 
  = \Av{{R_x(\BoldXi'+\XiR)}{R_x(\BoldXi')}}$ etc.
Note that the rightmost factor in Eq.~(\ref{gg//RR})
  is not the DC tensor itself 
  but its rearrangement,
  with the diagonal components exchanged.

Here we recall Eq.~(\ref{ChiR.L+T}) 
  to decompose the DC tensor 
  into the longitudinal and transverse correlations, 
  which reads as 
\begin{align}
  \ChiR
  &= 
  \begin{bmatrix}
  \Av{{R_x}{R_x}} & \Av{{R_x}{R_y}} \\
  \Av{{R_y}{R_x}} & \Av{{R_y}{R_y}}  
  \end{bmatrix}  
  \notag \\ 
  &= \Xl
  \begin{bmatrix} 
  {\CosSqPh}\! & \! {\SinPh\CosPh} \\ {\CosPh\SinPh}\! & \! {\SinSqPh} 
  \end{bmatrix}  
  \notag \\ &\qquad {}
  + \Xtr
  \begin{bmatrix} 
  {\SinSqPh}\! & \!{-\SinPh\CosPh} \\ {-\CosPh\SinPh}\! & \! {\CosSqPh} 
  \end{bmatrix}  
  \vphantom{{\TwoVect00}_{\frac00}}%
  \notag \\
  &= \frac{\Xl + \Xtr}{2} 
  \begin{bmatrix} 1 & 0 \\ 0 & 1 \end{bmatrix} 
  + \frac{\Xl - \Xtr}{2}
  \begin{bmatrix} 
  \cos{2\varphi_*}  &{~}\sin{2\varphi_*} \\ 
  \sin{2\varphi_*}  & {-\cos{2\varphi_*}} 
  \end{bmatrix}  
  \label{ChiR//2phi*}    
\end{align}
  in the polar coordinate system for the $\XiR$-space.
Taking notice of the last line in Eq.~(\ref{ChiR//2phi*}), 
  we define \begin{equation}
  f_\pm = \Xl \pm \Xtr 
  \label{f//RR}
\end{equation}
  for later convenience.
Then, with Eq.~(\ref{gg//RR}) in mind,
  we rearrange the matrix components in Eq.~(\ref{ChiR//2phi*})
  as 
\begin{multline}
  \begin{bmatrix}
  \Av{{R_y}{R_y}} & \Av{{R_x}{R_y}} \\
  \Av{{R_y}{R_x}} & \Av{{R_x}{R_x}}  
  \end{bmatrix}  
  \brX 
  = {\frac12}{f_+}
  \begin{bmatrix} 1 & 0 \\ 0 & 1 \end{bmatrix} 
  + {\frac12}{f_-}
  \begin{bmatrix} 
  {-\cos{2\varphi_*}}  &\sin{2\varphi_*} \\ 
  {~\sin{2\varphi_*}}  &\cos{2\varphi_*} 
  \end{bmatrix}  
  \relax,\hFil  
\end{multline}
  from which we obtain 
\begin{multline}
  \nabla_* \cdot 
  \begin{bmatrix}
  \Av{{R_y}{R_y}} & \Av{{R_x}{R_y}} \\
  \Av{{R_y}{R_x}} & \Av{{R_x}{R_x}}  
  \end{bmatrix}  
  \brX 
  = \frac{f_+'}{2\ell_0} \TwoVect{\CosPh}{\SinPh}
  + \ell_0^{-1} 
  \left( -\frac{f_-'}{2} + \frac{f_-}{\xi} \right)
  \TwoVect{~\cos{3\varphi_*}}{-\sin{3\varphi_*}}
  \label{d*RR},\hFil  
\end{multline} 
  where 
\begin{align*}
  \nabla_* 
  &= \ell_0^{-1} \TwoVect{\dXiAst }{\dHaAst } \notag \\ 
  &= \ell_0^{-1} \TwoVect{\CosPh}{\SinPh}\dXiR  
  + (\ell_0\xi_*)^{-1} \TwoVect{-\SinPh}{~\CosPh} \dPhi 
\end{align*}
  and $f_{\pm}' = \partial{f_\pm}/\partial{\xi_*}$.
Finally,  
  calculating the ($\xi_*$-space) divergence 
  of Eq.~(\ref{d*RR}),  
  we arrive at 
\begin{multline}
  \chi_\gm^{} 
  = -{\frac12}{\ell_0^{-2}}  
  {\xi_*^{-1}}\dd{}{\xi_*} \left(\,\xi_* f_+' \right)
  \brX  
  + \ell_0^{-2} \left(
  {\frac12} f_-'' 
  - \frac{5}{2}{\xi_*}^{-1} f_-' + 4\,{\xi_*}^{-2} f_-
  \right) \cos{4\varphi_*} 
  \label{gg//RR.polar}.\hFil 
\end{multline}
The $\varphi_*$-dependence of the expression 
  on the right-hand side Eq.~(\ref{gg//RR.polar}) 
  is consistent with Eq.~(\ref{gm.0+4}). 
We note 
  that the isotropic part is given in terms of $f_+$,  
  while the anisotropic part proportional to $\cos{4\varphi_*}$ 
  is determined by $f_-$. 
By comparing Eq.~(\ref{gg//RR.polar}) 
  with Eq.~(\ref{gm.0+4}) 
  and introducing linear operators of Euler--Cauchy type, 
\begin{subequations}%
\begin{align}
  \hat{L}_+ 
  &= -{\frac12}\ell_0^{-2}
  \left( \frac{\partial^2}{\partial\xi_*^2} + \xi_*^{-1} \dXiR \right)
  \label{L+},
  \\
  \hat{L}_- 
  &= \ell_0^{-2} \left(
  {\frac12} \frac{\partial^2}{\partial\xi_*^2} 
  - \frac{5}{2}{\xi_*}^{-1} \dd{}{\xi_*} + 4\,{\xi_*}^{-2} 
  \right) 
  \label{L-},
\end{align}
  we obtain 
\begin{equation}
  \chi_\gm^{(0)} = {\hat{L}_+ f_+} \relax, \quad 
  \chi_\gm^{(4)} = {\hat{L}_- f_-} \relax, 
  \label{chi04//L*f}%
\end{equation}%
\label{eqs:L}%
\end{subequations}%
  and Eq.~(\ref{gg//RR.polar}) is rewritten more concisely 
  as $\chi_\gm^{} 
  = {\hat{L}_+ f_+} + (\hat{L}_- f_-) \cos{4\varphi_*}$. 

\begin{figure}
  \centering 
  \includegraphics[clip,width=7.5cm]{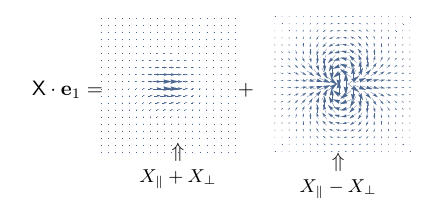} 
  \caption{\label{Fig:decomp}%
     Decomposition of the DC tensor into $f_\pm = \Xl\pm\Xtr$ 
     according to Eq.~(\ref{ChiR//2phi*});
     see Fig.~\protect\ref{Fig:RR.2D}~(b).
  }
\end{figure}

In comparison to Eq.~(\ref{gm.0+4}) for $\chi_\gm$,  
  which is understood simply 
  as a special case of Fourier decomposition with regard to $\varphi_*$, 
  the decomposition of the DC tensor into $f_+$ and $f_-$ 
  in the form of Eq.~(\ref{ChiR//2phi*}) 
  may seem more difficult to conceive.
Pictorially, 
  this decomposition can be illustrated as in Fig.~\ref{Fig:decomp}, 
  to be compared with Fig.~\ref{Fig:RR.2D}.
The component given by $f_+ = \Xl + \Xtr$ 
  represents correlated motion 
  in which all the neighboring particles are dragged 
  in the same direction, 
  while the so-called back\-flow pattern 
  is represented by $f_- = \Xl - \Xtr$ as a vortex dipole.

\subsection{Possible nonequivalence between DC and SC}
\label{subsec:L+}

With the relation in Eqs.~(\ref{eqs:L}) 
  we can calculate the SC from the DC, 
  but this does not necessarily mean that SC and DC are equivalent.
The problem is whether the relation is invertible, 
  making it possible to obtain the DC from the SC. 
Intuitively speaking, 
  since differentiation is involved in the operators $\hat{L}_\pm$, 
  some information in DC is likely to be missing from SC. 

The invertibility 
  of the linear operators $\hat{L}_\pm$ in Eq.~(\ref{eqs:L}) 
  can be discussed by checking their null spaces.
The null space of $\hat{L}_+$, 
  also known as the kernel of the linear operator 
  \cite{Rynne.Book2007}, 
  is the set of all the solutions $h_+$ to the homogeneous equation
\begin{equation}
  \hat{L}_+ h_+ = 0
  \label{LH.L+}, 
\end{equation}
  with the domain of the function
  taken as some physically appropriate range of $\xi_*$. 
Noticing that $\hat{L}_+$ is essentially the radial part 
  of the Laplacian operator in the $\XiR$-space, 
  we can readily find $h_+ = A + B\log\xi_*$ (with $A$ and $B$ 
  denoting arbitrary constants); 
  in other words, 
  the null space is spanned by $\{1,\log\xi_*\}$. 
If $f_+ \sim h_+$,  
  it is mapped to zero 
  by $\hat{L}_+$ in Eqs.~(\ref{eqs:L}) 
  and therefore the information of $f_+$ 
  is lost from $\chi_\gm^{(0)}$. 
This makes the DC and the SC nonequivalent.

Analogously,  
  the general solution to the linear homogeneous equation
\begin{equation}
  \hat{L}_- h_- = 0
  \label{LH.L-}
\end{equation}
  comprises the null space of $\hat{L}_-$, 
  which is spanned by $\{ \xi_*^2, \xi_*^4 \}$.  
If $f_-$ happens to fall into the null space of $\hat{L}_-$, 
  the information of $f_-$ drops out of $\chi_\gm^{(4)}$. 

Thus we have derived
  analytical relations involving the SCs,
  which allows us to discuss the possibility
  that the DCs and the SCs are not equivalent
  due to the null space of $\hat{L}_\pm$.
Now let us proceed to numerical study of these analytical results,
  beginning with specification of the particle system 
  in the next section.



\section{Specification of the particle system}
\label{sec:system}%

Let us specify the particle system as a model liquid,  
  for which we calculate the DC and the SC numerically.
We consider 
  a system consisting of $N$ Brownian particles (disks)
  in a 2D periodic box of the size $L^2$, 
  which is basically the same system  
  as in our previous work \cite{Ooshida.PRE94}.
The position vectors of the particles,
  denoted by $\mb{r}_i$ with $i = 1, 2, \ldots, N$, 
  are governed 
  by the over\-damped Langevin equation  
\begin{equation}
  \mu \dot{\mb{r}}_i  
   = - \dd{}{\mb{r}_i} \sum_{j<k} V_{jk}
     + \mu \mb{f}_i(t)
  \label{Langevin-.r},
\end{equation}
  where 
  $\mu$ is the drag coefficient, 
  and $\mu\mb{f}_i(t)$ is the thermal fluctuation term
  with the temperature $T$,
  corresponding to the bare diffusivity $D = \kT/\mu$
  and prescribed as a Gaussian random forcing 
  with zero mean and the variance
\[
  \Av{\mb{f}_i(t) \otimes \mb{f}_j(t')}
  = 2D \delta_{ij} \delta(t-t') \openone
  \relax. %
\]
As the interaction potential $V_{jk}$ 
  between the $j$-th and $k$-th particles 
  (separated by the relative position vector 
  $\mb{r}_{jk}$), 
  we adopt the harmonic repulsive potential,
\[
  V_{jk} = 
  \begin{cases}
  \Vmax \left( 1 - \dfrac{\abs{\mb{r}_{jk}}}{\sigma} \right)^2
  \relax & (\abs{\mb{r}_{jk}} < \sigma) \cr 
  0      & (\text{otherwise})
  \end{cases}
\]%
  with very large barrier $\Vmax$, 
  nearly equivalent to the hardcore interaction
  with diameter $\sigma$ (the particles are mono\-disperse).
The inertia 
  is completely ignored.

Note that here we have adopted a system of mono\-disperse particles, 
  which makes it simpler to specify the system.
This is allowed 
  as we are interested in the validity of the solid-based approach 
  to only slightly glassy liquids.

We prepare the system
  to be in a statistically steady and homogeneous state
  in equilibrium at the temperature $T$.
The mean density is $\rho_0 = N/L^2$,
  with which we define $\ell_0 = 1/\sqrt{\rho_0}$ 
  as a length scale 
  that represents the typical inter\-particle distance.
The area fraction is  
  $\phiA = (\pi/4)\sigma^2 \rho_0 = (\pi/4)(\sigma/\ell_0)^2$.


\section{Numerical results}
\label{sec:num}


We performed numerical simulation 
  of the system of Brownian disks specified in Sec.~\ref{sec:system}.
The system contains $N = 4000$ particles,
  and $L$ is adjusted so as to give three values of the area fraction: 
  $\phiA = 0.50$, $0.60$ and $0.70$.
(For higher densities,
  poly\-disperse particles would be needed to avoid crystallization, 
  which is out of the scope of the present work.)
To mimic the hardcore interaction, we chose 
  $\Vmax = 50$   for $\phiA = 0.50$,
  $\Vmax = 500$   for $\phiA = 0.60$, and
  $\Vmax = 5000$ for $\phiA = 0.70$ in units of $\kT$;
  the time step for numerical integration, $\Delta{t}$,
  is chosen so as to satisfy 
  $(\Delta{t}/\tau_0)(\kT/\Vmax) = 0.1$ where $\tau_0 = \sigma^2/D$.

In what follows,  
  reference scales for nondimensionalization, 
  such as $\sigma$ and $\ell_0$,  
  will be shown explicitly as a rule.
In referring to the time interval $\tD$, however, 
  we will make an exception:  
  we will write simply $t = 20$, for example,  
  instead of $\tD/\tau_0 = 20$.


\subsection{Detectability of DC and SC in only slightly glassy liquids}

\begin{table}
  \centering
  \caption{\label{Tab:S}%
    Long\-wave limiting values of the static structure factor, $S$, 
    and the structural relaxation time, $\tauA$, 
    obtained from 2D particle simulation data.
    The values of $\ell_0/\sigma$ are also included in the table.
  }%
  \begin{tabular}[t]{c|lll}
  $\phiA$         & 0.50  & 0.60  & 0.70  \\ \hline
  $S$             & 0.16~ & 0.055 & 0.025 \\
  $\tauA/\tau_0$  & 0.032 & 0.050 & 0.159 \\ %
  $\ell_0/\sigma$ & 1.253 & 1.144 & 1.059 
  \end{tabular}
\end{table}

\begin{figure}[b]
\centering
\raisebox{5.7cm}{(a)}\!\includegraphics[clip,height=6.0cm]{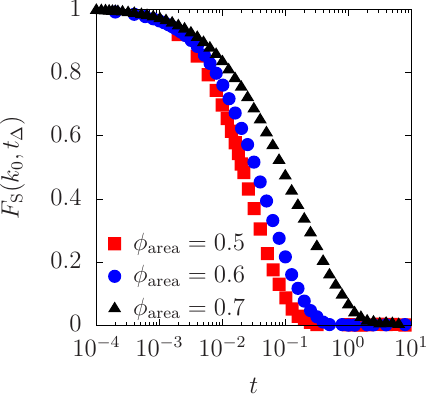}\quad 
\raisebox{5.7cm}{(b)}\!\includegraphics[clip,height=6.0cm]{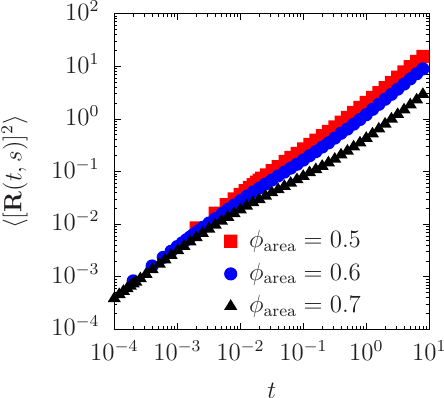}
\caption{\label{Fig:Fs}%
  Time dependence of well-established statistical quantities 
  based on single particle displacement, 
  computed for three different values of the area fraction 
  ($\phiA = 0.5$, $0.6$, and $0.7$).
  Note that the axis label $t$ actually stands for $\tD/\tau_0$.
  (a) The self part of the intermediate scattering function, 
  $\Fs(k,\tD)$, at $k = k_0 \;(=2\pi/\sigma)$.
  (b) The MSD.
}
\end{figure}

We start with calculating 
  the quantities tabulated in Table~\ref{Tab:S}
  for the range of $\phiA$ under consideration,
  where $S$ denotes 
  the long\-wave limiting values of the static structure factor.
The $\alpha$ relaxation time, $\tauA$, 
  is determined by the condition that $\Fs(k_0,\tauA) = 1/e$, 
  with $\Fs$ denoting  
  the self part of the intermediate scattering function, 
\begin{equation}
  \Fs(k,\tD) 
  = \Av{{\frac{1}{N}}\sum_i%
  \exp\left[{\IK\cdot\mb{R}_i(t,s)}\right]}
  \quad (t=s+\tD)
  \label{Fs}, 
\end{equation}  
  and $k_0 = 2\pi/\sigma$ \cite{Yamamoto.PRE58,Kob.PRE52}.
Plots of $\Fs(k_0,\tD)$
  are shown in Fig.~\ref{Fig:Fs}
  along with the MSD, 
  $\Av{\left[{\mb{R}(t,s)}\right]^2}$.
For $\tD\gg\tauA$,
  the MSD grows asymptotically in proportion to $\tD$, 
  demonstrating 
  that the separate ``steps'' in Eq.~(\ref{R//sum.step}) 
  are uncorrelated.

Noticing the modesty of $\tauA$ in Table~\ref{Tab:S}, 
  which indicates that the liquid is only slightly glassy, 
  we recall the appreciable presence of DCs even in such cases, 
  as was reported in our previous work \cite{Ooshida.PRE94}
  and reaffirmed in Fig.~\ref{Fig:RR.2D} of the present article.  
In regard to the displacement $\mb{R}$ 
  in the case of such modest $\tauA$,
  it should be noted 
  that the time averaging implied in Eq.~(\ref{R=}) 
  does not suffice by itself 
  to get rid of the noisiness from displacement-based statistical quantities. 
Even in such cases, 
  with the aid of ensemble averaging,
  the statistical procedure in Appendix~\ref{app:DC} 
  reveals appreciable presence of DCs, 
  as we have seen in Fig.~\ref{Fig:RR.2D} of the present article
  and in Figs.~5, 6, and 8 of Ref.~\cite{Ooshida.PRE94}, 
  not only for $t=0.50$ but also for $t=15.9$. 
In other words, 
  displacements are definitely correlated 
  in such an only slightly glassy liquid ($\phiA = 0.5$).

  
\begin{figure}
\centering
\raisebox{6.0cm}{(a)}\includegraphics[clip,width=7.0cm]{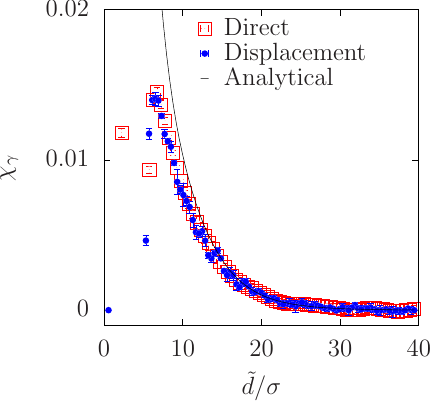}\quad
\raisebox{6.0cm}{(b)}\includegraphics[clip,width=7.0cm]{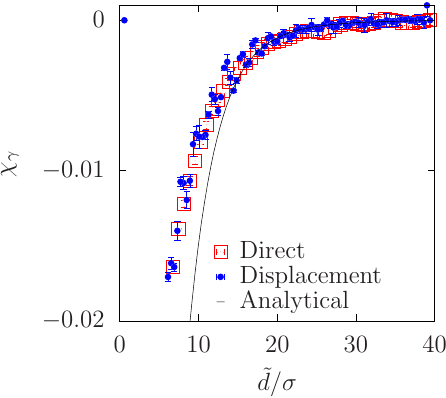}
\caption{\label{Fig:cmp3}%
    Comparison among three distinct evaluations of the SC
  for $t = 20$ and $\phiA = 0.5$.
  The two panels correspond to different directions of $\di$: 
  (a) $\varphi_* = 0$, and (b) $\varphi_* = \pi/4$.
  The red squares indicate 
  direct computation from the particle simulation data 
  with Eq.~(\ref{chiGm=}),
  while indirect evaluation from the DC 
  by way of Eq.~(\ref{gg//RR.polar})
  is plotted with blue solid circles,
  and analytical expression in Eq.~(\ref{chiGm//th}) 
  with $\muR = 0.1$ 
  is shown with a thin solid line in each panel.%
}
\end{figure}

Let us proceed 
  to the calculation of the SC. 
The angular dependence of SC, 
  involving $\cos{4\varphi_*}$ 
  as in Eqs.~(\ref{gm.0+4}) and (\ref{gg//RR.polar}), 
  is almost self-evident:
  the isotropy of the system 
  implies that the correlation of $\gm$, 
  which is a component of a second order tensor, 
  should have the fourfold angular symmetry. 
Instead of the angular dependence, 
  we should rather focus on the dependence 
  on the distance 
  $\di[\relax]$, 
  with the angle $\varphi_*$ fixed.

Since the SCs and the DCs 
  are theoretically predicted 
  to be related by Eq.~(\ref{gg//RR.polar}), 
  we can expect presence of SC 
  in cases in which DC is detectable.
In order to validate this qualitative expectation,  
  along with the quantitative relation in Eq.~(\ref{gg//RR.polar}),
  we computed $\chi_\gm$ from the simulation data, 
  using the definition in Eq.~(\ref{chiGm=}) 
  and following the numerical procedure 
  in the supplemental material 
  of Illing \textit{et al.}~\cite{Illing.PRL117}.

The values of SC thus calculated 
  by direct usage of Eq.~(\ref{chiGm=}) 
  are plotted in Fig.~\ref{Fig:cmp3}.
For comparison, 
  we have also included the values 
  calculated indirectly by way of Eq.~(\ref{gg//RR.polar}) 
  from the DCs.
It is evident from Fig.~\ref{Fig:cmp3}
  that the values of SC are in reasonable agreement,  
  except for the shorter range with $\di[\relax] < 7\sigma$ 
  for which it is difficult 
  to obtain reliable values of SC.
Thus, 
  at least in the case of Fig.~\ref{Fig:cmp3}
  and for $\di[\relax] > 7\sigma$, 
  we have demonstrated 
  the detectability of the SC 
  as well as the validity of Eq.~(\ref{gg//RR.polar}) 
  that relates the DC and the SC. 

The solid lines in Fig.~\ref{Fig:cmp3} 
  represent analytical curves 
  based on an approximate expression for SC,
  to be given later 
  as Eq.~(\ref{chiGm//th}) in Subsec.~\ref{subsec:elastic}.
The behavior of the curve, 
  decaying to zero as $\di[\relax]$ increases, 
  is qualitatively consistent with the numerical results,  
  though some disagreement for small $\di[\relax]$ is also visible.

\subsection{Indication of nonequivalence between SC and DC}

Having seen the validity of Eq.~(\ref{gg//RR.polar}) 
  relating the DC and the SC, 
  it is natural to ask 
  whether the two kinds of correlations are equivalent.
We start answering this question
  with comparative study of their dependences on $\phiA$.

\begin{figure}
  \centering
  \raisebox{5.5cm}{(a)}\includegraphics[clip,height=6.0cm]{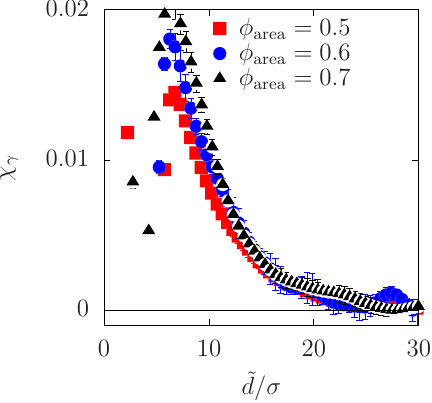}\quad
  \raisebox{5.5cm}{(b)}\includegraphics[clip,height=6.0cm]{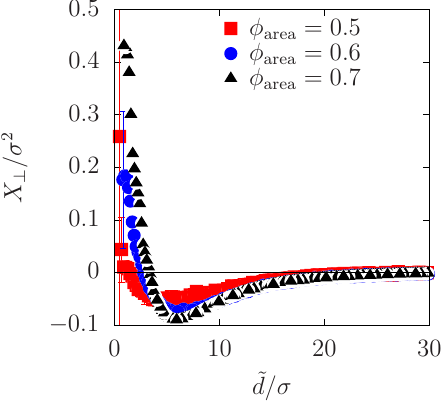}
  \caption{\label{Fig:phiA}%
    Dependence of the correlations 
  on the area fraction ($\phiA = 0.5$, $0.6$, and $0.7$).
  (a) SC with $\varphi_* = 0$.
  (b) Transverse DC.
}
\end{figure}

In Fig.~\ref{Fig:phiA}(a),  
  the SCs (at $\varphi_* = 0$) 
  for $\phiA = 0.50$, $0.60$ and $0.70$ 
  are compared, 
  while the corresponding plots of $\Xtr$, the transverse DC, 
  is shown in Fig.~\ref{Fig:phiA}(b).
Evidently, 
  $\chi_\gm$ in Fig.~\ref{Fig:phiA}(a) 
  exhibits weaker dependence on $\phiA$
  than $\Xtr$ in Fig.~\ref{Fig:phiA}(b).  
In other words, 
  DCs are more sensitive to $\phiA$ than SCs;
  this difference suggests 
  that DCs and SCs are not equivalent.

By closer observation on Fig.~\ref{Fig:phiA}, 
  we find a significant difference 
  in the shorter-distance range with $\di[\relax] < 7\sigma$.
As was noticed in the previous subsection,    
  it is difficult to obtain reliable values of SC
  in the shorter-distance range.
Contrastively, 
  the values of DCs seem still reliable in this range,
  as the numerical values appear to behave consistently 
  throughout the almost entire range of $\di[\relax] > \ell_0$.  
We also notice 
  that, in the cases of $\phiA=0.60$ and $0.70$, 
  there is a significant increase in $\ChiR$ 
  as $\di[\relax]$ approaches zero.
No counterpart of this $\phiA$-dependent increase in DC 
  seems to be recognizable in SC. 

These observations on the behavior of DC and SC 
  can be explained 
  in terms of the null space 
  of the operator $\hat{L}_+$ given in Eq.~(\ref{L+}). 
As was discussed in Subsec.~\ref{subsec:L+}, 
  the mapping from $f_\pm$ to $\chi_\gm$ 
  is not invertible in general, 
  as long as $f_\pm$ may fall into the null space of $\hat{L}_\pm$.
This makes the DC and the SC non\-equivalent.
  
Let us discuss 
  $\hat{L}_+$ and $\hat{L}_-$ separately.
The null space of $\hat{L}_-$
  consists of the general solution to Eq.~(\ref{LH.L-}),  
  spanned by $\{ \xi_*^2, \xi_*^4 \}$, 
  whose relevance can be ruled out 
  by the observation 
  that the DCs do not exhibit such a strong divergence 
  for $\xi_* = \di[\relax]/\ell_0 \to +\infty$.
Contrastively, 
  from the solutions to Eq.~(\ref{LH.L+}), 
  we find that $\ln\xi_*$ belongs to the null space of $\hat{L}_+$, 
  which seems to give a consistent explanation 
  of the numerical observations.
If $f_+$ behaves as $\ln\xi_*$ 
  in some shorter-distance range of $\xi_*$,
  the components of the DC tensor 
  also diverge as
\[ \Xl \sim \Xtr \sim \ln\xi_*, \] 
  while the information is lost from $\chi_\gm$,  
  because $\hat{L}_+ \ln\xi_* = 0$.
Thus the numerical observations 
  indicating non\-equivalence between SC and DC 
  can be explained 
  if the DC behaves logarithmically for small $\xi_*$.

\subsection{Logarithmic behavior of DCs}
\label{subsec:log.data}

\begin{figure}
  \centering
  \raisebox{5.5cm}{(a)}\includegraphics[clip,height=6.0cm]{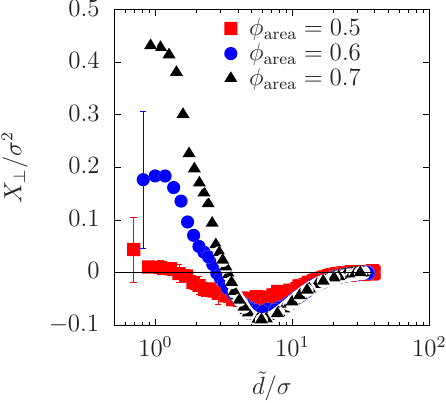}\quad
  \raisebox{5.5cm}{(b)}\includegraphics[clip,height=6.0cm]{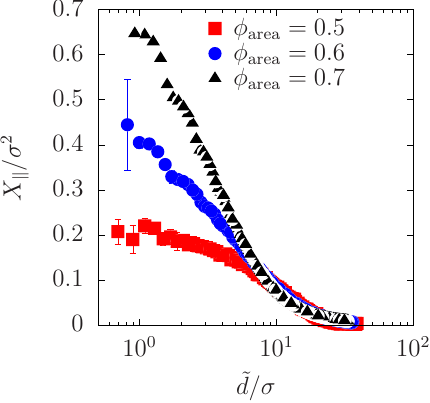}
  \caption{\label{Fig:ChiR.log}%
    DCs plotted against the spatial separation $\di[\relax]$,
    with the $\di[\relax]$-axis scaled logarithmically.
    In both panels, DCs computed at $t=20$ 
    for $\phiA = 0.5$, $0.6$, and $0.7$ are shown.
    (a) Transverse DCs.
    (b) Longitudinal DCs.
  }
\end{figure}

To establish 
  the explanation of the non\-equivalence between SC and DC 
  due to the null space of $\hat{L}_+$, 
  now let us demonstrate 
  the presence of logarithmic behavior in DC. 
This is demonstrated 
  by plotting the components of $\ChiR$ 
  against the logarithm of $\xi_* = \di[\relax]/\ell_0$
  and then finding the range of $\xi_*$ 
  over which the plots form straight lines. 

In Fig.~\ref{Fig:ChiR.log}
  we have such a semi\-logarithmic plot (of linear--log type).
A narrow but recognizable range of straight-line behavior
  is present.
We also notice 
  that the slope of the plots for $\Xl$ 
  is nearly the same as that for $\Xtr$.
Besides, 
  the slope becomes steeper as $\phiA$ is increased.

\subsection{Time dependence of SC and DC: diffusive scaling}

\begin{figure}
\centering
\raisebox{5.0cm}{(a)}~\includegraphics[clip,width=6.0cm]{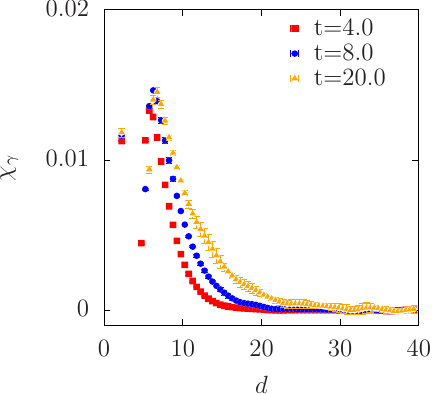}\quad
\raisebox{5.0cm}{(b)}~\includegraphics[clip,width=6.0cm]{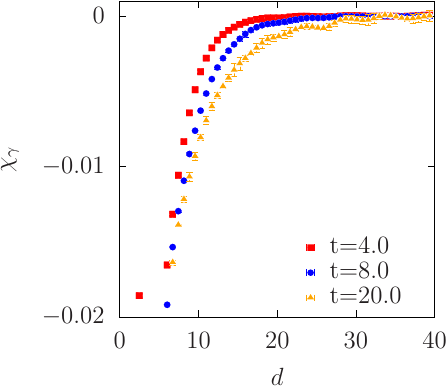}
\caption{\label{Fig:gg@t}%
  SCs versus $\di[\relax]$ for three different time interval,
  computed for (a) $\varphi_* = 0$ and (b) $\varphi_* = \pi/4$.
}
\end{figure}

It seems rather surprising 
  that correlations at finite $\xi_*$, such as DC and SC, 
  persists for $\tD\gg\tauA$.
Some authors 
  proposed to interpret this persistence 
  as an accumulated effect of many events with shorter correlation time;
  this was proposed by Doliwa and Heuer \cite{Doliwa.PRE61} 
  in regard to DC, 
  and later by Chattoraj and {\Lemaitre} \cite{Chattoraj.PRL111} 
  for SC. 
This interpretation 
  leads to the prediction 
  that $\ChiR(\di,\tD)$ and $\chi_\gm(\di,\tD)$ 
  are proportional to $\tD$ (with $\di$ fixed), 
  as reviewed at the end of Subsec.~\ref{subsec:DC}.

This prediction, however, 
  is inconsistent with the numerical results 
  shown in Fig.~\ref{Fig:gg@t},
  in which plots of SC are compared 
  for three different values of $\tD$. 
The SCs (with $\di$ fixed) 
  do not grow in proportion to $\tD$,  
  but rather seem to exhibit some different type of $\tD$-dependence through a similarity variable.

\begin{figure}
  \centering
  \raisebox{5.5cm}{(a)}\includegraphics[clip,width=6.0cm]{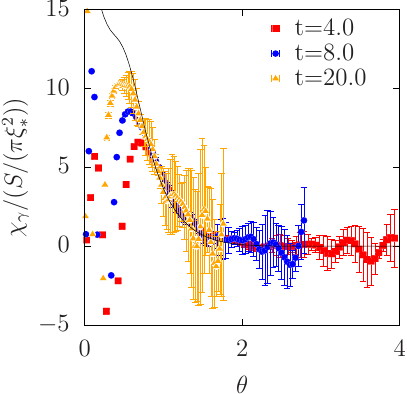}\quad 
  \raisebox{5.5cm}{(b)}\includegraphics[clip,width=6.0cm]{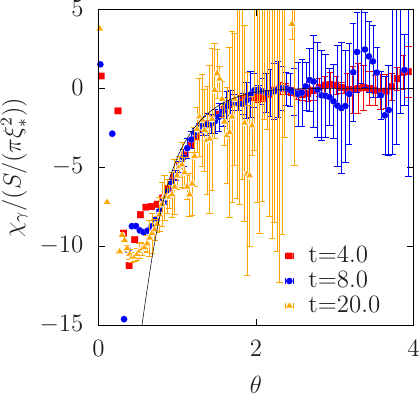}
  \caption{\label{Fig:sim}%
    Test for expressibility of SC in terms of the similarity variable 
    $\vartheta=\xi_*/(2\sqrt{\Dc_*{t}})$.
    The same data as in Fig.~\protect\ref{Fig:gg@t} are used
    and shown to collapse into single curves
    for (a) $\varphi_* = 0$ and (b) $\varphi_* = \pi/4$.
  }
\end{figure}

In our previous work \cite{Ooshida.PRE94} on DC 
  (mainly with $\phiA = 0.5$),  
  we found that $\Xl$ and $\Xtr$ are expressible 
  in terms of a similarity variable,
\begin{equation}
  \vartheta 
  {} = \frac{\di[\relax]}{2\sqrt{{\Dc}\tD\mathstrut}}  
  = \frac{\xi_*}{2\sqrt{{\Dc_*}\tD\mathstrut}}
  \label{sim.theta}, 
\end{equation}
  including the diffusive length scale 
  $2\sqrt{{\Dc}\tD}$ where $\Dc = D/S$ and $\Dc_* = \Dc/\ell_0^2$; 
  see Eq.~(2.10) and Eq.~(4.38) in Ref.~\cite{Ooshida.PRE94}
  about this length scale.
The result suggests 
  that the cages are nested 
  to form a self-similar structure in the space--time.

Since the SCs are related with the DCs 
  by Eq.~(\ref{gg//RR.polar}), 
  we expect that the SCs are also expressible 
  in terms of the same similarity variable $\vartheta$.
This is verified 
  by taking the data of SC in Fig.~\ref{Fig:gg@t} 
  and re\-plotting them against $\vartheta$. 
As a result, 
  the plots for different $\tD$ 
  are seen to collapse into a single curve, 
  as is shown in Fig.~\ref{Fig:sim}.
Note that the error bars in Fig.~\ref{Fig:gg@t}
          are magnified in Fig.~\ref{Fig:sim} by the factor of $\xi^2$
          and therefore appears to be large.

\begin{figure}
  \raisebox{5.0cm}{(a)}\includegraphics[clip,width=7.0cm]{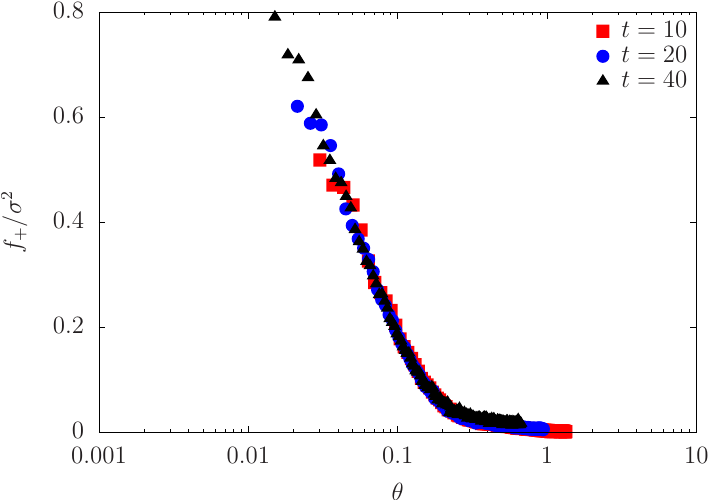}\quad
  \raisebox{5.0cm}{(b)}\includegraphics[clip,width=7.0cm]{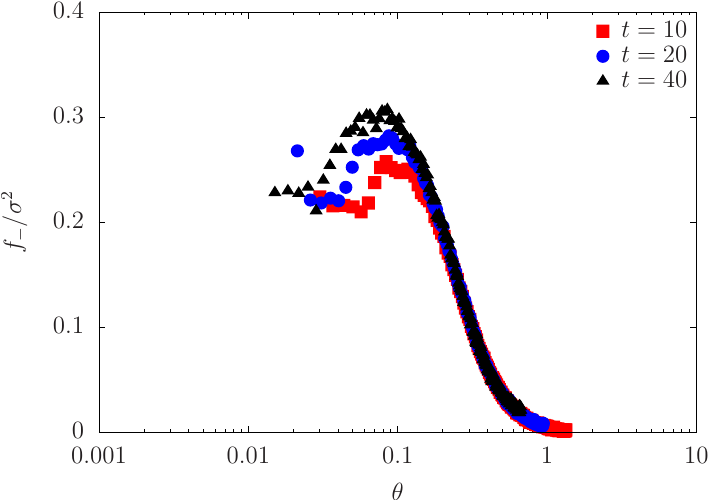}
  \caption{\label{Fig:f//th}%
     $f_\pm$ defined by Eq.~(\ref{f//RR}), 
     computed for $\phiA = 0.60$
     and plotted against $\vartheta$.
  }
\end{figure}
\begin{figure}
  \centering
  \includegraphics[clip,width=7.0cm]{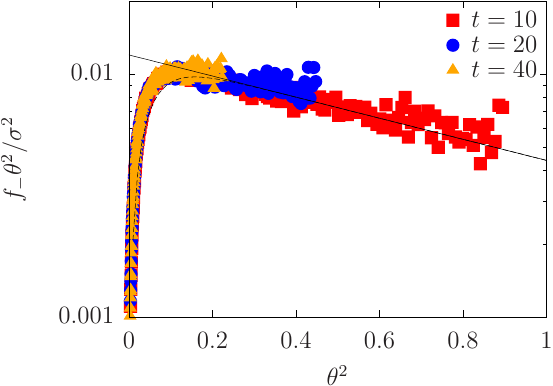}
  \caption{\label{Fig:th2*f-}%
          Replot of $f_-$ 
     based on the same data as in Fig.~\protect\ref{Fig:f//th}(b).
     The curves collapse onto a single straight line 
     by plotting $\vartheta^2 f_-$ against $\vartheta^2$,
     with the vertical axis in the logarithmic scale.
  }
\end{figure}

Finally, let us examine
  $f_\pm$ defined by Eq.~(\ref{f//RR}).
They are plotted against the similarity variable $\vartheta$
  in Fig.~\ref{Fig:f//th},
  with the $\vartheta$-axis in logarithmic scale.
The data of $f_+$ for different time intervals 
  are seen to collapse on a single curve in Fig.~\ref{Fig:f//th}(a),
  and the curve 
  is close to a straight line (indicating logarithmic behavior)
  for $\vartheta < 0.1$,
  i.e.\ 
  for distances much shorter than the diffusive length scale.
Thus $f_+$ is shown to behave logarithmically 
  at shorter distances.
Contrastively, 
  as is seen in Fig.~\ref{Fig:f//th}(b),
  the behavior of $f_-$ in the range of $\vartheta < 0.1$ 
  is not logarithmic at all.
For $0.3 < \vartheta < 1$,
  the data of $f_-$ are seen to collapse on a single curve,
  but the semi-logarithmic plot in Fig.~\ref{Fig:f//th}
  fails to make this curve a straight line.
A different kind of plot gives a better explanation for $f_-$: 
  as is shown in Fig.~\ref{Fig:th2*f-},
  the curve is straightened 
  by plotting $\vartheta^2 f_-$ in logarithmic scale 
  against $\vartheta^2$ in linear scale.
This implies that $f_-$ behaves as $\vartheta^{-2} e^{-\vartheta^2}$ 
  at longer distances ($\vartheta \sim 1$).

\section{Discussion}
\label{sec:discussion}

For the generic framework proposed in Sec.~\ref{sec:relation}
  to relate the SCs with other correlations, 
  we verified numerically, in the previous section, 
  that the SC can be obtained 
  from the components of the DC tensor. 
We have also shown 
  that the computed DCs behave logarithmically 
  on shorter lengthscales.
The slope or the amplitude of the logarithmic part 
  depends on the area fraction $\phiA$, 
  but the corresponding $\phiA$-dependence is missing from the SCs, 
  as was confirmed numerically 
  in Sec.~\ref{sec:num} with Fig.~\ref{Fig:phiA}.

Now let us discuss 
  what kind of information can be read 
  from the logarithmic behavior of the DC.
We start with a rough but simple modeling 
  by fluctuating elastic media, 
  which provides 
  specific expressions of $\Cd^\Delta$ and $\Cr^\Delta$ 
  as inputs into the Alexander--Pincus formula (\ref{eqs:I.AP2})
  to calculate the DCs concretely.
Subsequently, 
  we will extend our discussion 
  to a wider class of models for $C_a^\Delta$,  
  showing that the logarithmic regime of the DC 
  can be caused 
  by a certain kind of ``caged'' behavior of $C_a^\Delta$. 

\subsection{Elastic modeling}
\label{subsec:elastic}

The formulae in Subsec.~\ref{subsec:formula}, 
  relating $C_a^\Delta$ to the DC and the SC, 
  are generic 
  in the sense that dynamics of $\psi_a$ are not specified.
Although the dynamics for the particles 
  are given by the Langevin equation (\ref{Langevin-.r}), 
  it is not obvious 
  how the dynamics are 
  projected onto those of $\psi_a$ 
  to yield manageable expression of $C_a^\Delta$.
To proceed further, 
  here we assume 
  some approximate dynamics that give 
\begin{subequations}
\begin{align}
  \Cd^\Delta(\K,t,s) = \Cd^\Delta(k,\tD)
  &= S\,\left( 1 - e^{-\Dc_*\K^2 \tD} \right)
  \label{Cd.Delta.exp},
  \\
  \Cr^\Delta(\K,t,s) = \Cr^\Delta(k,\tD)
  &= \frac{S}{\muR}\,\left( 1 - e^{-\muR\Dc_*\K^2 \tD} \right)
  \label{Cr.Delta.exp},
\end{align}%
\label{eqs:Corr.Delta.exp}%
\end{subequations}%
  as is proposed 
  in the discussing section of Ref.~\cite{Ooshida.PRE94};
  here $S$ and $\muR$ are positive constants 
  \cite{note.muR}, 
  and $\Dc_* = \Dc/\ell_0^2 = (D/S)/{\ell_0^2}$.
To be consistent 
  with the known dynamics of $\rho(\mb{r},t)$, 
  the constant $S$ is chosen to be equal 
  to the long\-wave limiting value of the static structure factor
  (see Table~\ref{Tab:S}).

The correlations proposed in Eq.~(\ref{eqs:Corr.Delta.exp}) 
  are understood as a result of elastic modeling
  for the dynamics of the deformation gradient field, 
  which may be formulated 
  as Langevin equations in the following form:
\begin{subequations}
\begin{align}
  (\dt + \Dc_*    \,\mb{k}^2)\psiD(\mb{k},t) &= \fd(\mb{k},t)
  \label{dt.psiD.lin},
  \\
  (\dt + \muR\Dc_*\,\mb{k}^2)\psiR(\mb{k},t) &= \fr(\mb{k},t)
  \label{dt.psiR.muR},
\end{align}%
\label{eqs:dt.psi.muR}%
\end{subequations}
  where $\fd$ and $\fr$ are thermal fluctuation terms 
  with the temperature $T$.
The term with the coefficient $\Dc_* \propto \kT/S$ 
  in Eq.~(\ref{dt.psiD.lin})
  represents ``restoring force'' exerted upon $\psiD$ 
  by the elastic medium with the bulk modulus $\kT/S$.
We note, parenthetically, 
  that a 1D version of Eq.~(\ref{dt.psiD.lin})
  has often been used to describe 
  continuum dynamics of a chain of colloidal particles
  in a channel \cite{Lizana.PRE81,Ooshida.BRL11,Ooshida.JPCM30}.

In parallel to the term with $\Dc_*$ in Eq.~(\ref{dt.psiD.lin}), 
  the presence of the term with $\muR\Dc_*$ 
  in Eq.~(\ref{dt.psiR.muR})
  means introducing the shear modulus $\muR\kT/S$
  into the model.
While this is a natural modeling of DCs 
  in the case of glass solids with plateau modulus
  or with idealization of $\tauA\to\infty$ 
  \cite{Toninelli.PRE71,Klix.PRL109,Flenner.PRL114},
  it may seem questionable 
  to what extent the elastic modeling is applicable 
  to the case of glassy liquids with modest $\tauA$.
Here we regard the elastic modeling as a convenient starting point 
  for searching more sophisticated description of $C_a^\Delta$ 
  with a wider range of applicability,
  involving solid-like elasticity at some scale.
Recent experiments on confined 
  liquid glyserol \cite{Kume.PhF33,Kume.SciRep10} %
  and polypropylene glycol \cite{Kume.SciRep10}
  may support relevance of such solid-like elasticity 
  in liquids far away from glass transition.

Given the numerical data of the DCs,
  we focus our attention 
  to the ratio of the coefficients $S$ and $S/\muR$ 
  in Eqs.~(\ref{eqs:Corr.Delta.exp}).
In principle, 
  $\muR$ can be computed from the data in the Fourier space
  as the ratio of the saturation values 
  of $\Cd^\Delta$ and $\Cr^\Delta$:
\begin{equation}
  \muR = \lim_{\tD\to\infty} 
  \frac{\Cd^\Delta(k,\tD)}{\Cr^\Delta(k,\tD)}
  \label{muR=Cd/Cr}.
\end{equation}  
In the case of solidified glass,
  $\muR$ represents 
  the ratio of the shear modulus to the bulk modulus, 
  which certainly makes sense for any $k$ small enough 
  to justify elastic modeling 
  and with $\tD\to\infty$ understood 
  within the range of $\tD\ll\tauA$ \cite{Klix.PRL109,Flenner.PRL114}.
Here we relax this restriction on $\tD$, however, 
  expecting 
  that Eq.~(\ref{eqs:Corr.Delta.exp}) can be still valid 
  for $\tD > \tauA$ in some range of $k$ 
  and may serve as a useful starting point for discussion.

With $\Cd^\Delta$ and $\Cr^\Delta$ 
  given in Eq.~(\ref{eqs:Corr.Delta.exp}), 
  we can calculate the DC and the SC analytically, 
  using the formulae in Subsec.~\ref{subsec:formula}.

Let us begin with the Alexander--Pincus formula (\ref{eqs:I.AP2}) 
  for the DC tensor.
Before staring the calculation, 
  we note that the mathematical procedure 
  is somewhat simplified by utilizing  
\begin{equation}
  \dd{}{\tD} \left( {\Cd^\Delta},\; 
  \Cr^\Delta \right)
  = \frac{Dk^2}{\ell_0^2}\,
  \left( {e^{-\Dc_*{k^2}\tD}}, \; {e^{-\muR\Dc_*{k^2}\tD}} \right)
  \label{dt.Corr.Delta.exp}.
\end{equation}%
Combining Eq.~(\ref{dt.Corr.Delta.exp})
  with the Alexander--Pincus formula 
  in the polar coordinate form in Eq.~(\ref{eqs:AP2.polar}), 
  we have 
\begin{widetext}%
\begin{subequations}
\begin{align}
  & \dd{\Id }{\tD} 
  = \frac{{\pi}D}{\ell_0^2}\int_0^\infty 
  {e^{-\Dc_*{k^2}\tD}} \left\{
  \begin{bmatrix} 1 & 0 \\ 0 & 1 \end{bmatrix}
  J_0(k\xi_*) 
  -  
  \begin{bmatrix}
  \cos{2\varphi_*}  & \,\sin{2\varphi_*} \\
  \sin{2\varphi_*}  & - \cos{2\varphi_*} 
  \end{bmatrix}
  J_2(k\xi_*)
  \right\}
  k\D{k}
  \label{Id//dC.elastic},
  \\
  & \dd{\Ir }{\tD}
  = \frac{{\pi}D}{\ell_0^2}\int_0^\infty 
  {e^{-\muR\Dc_*{k^2}\tD}} \left\{
  \begin{bmatrix} 1 & 0 \\ 0 & 1 \end{bmatrix}
  J_0(k\xi_*) 
  + 
  \begin{bmatrix}
  \cos{2\varphi_*}  & \,\sin{2\varphi_*} \\
  \sin{2\varphi_*}  & - \cos{2\varphi_*} 
  \end{bmatrix}
  J_2(k\xi_*)
  \right\}
  k\D{k}
  \label{Ir//dC.elastic}.
\end{align}%
\label{eqs:AP2//elastic}%
\end{subequations}%
\end{widetext}%
By evaluating the wavenumber integrals 
  according to Appendix~\ref{app:intK.Bessel.gauss}
  and then calculating the antiderivatives with regard to $\tD$, 
  we obtain $\Id$ and $\Ir$ 
  as functions of $\xi_*$, $\varphi_*$ and $\tD$.
The result turns out to be expressible
  in terms of the similarity variable 
  $\vartheta$ in Eq.~(\ref{sim.theta});
  then, substituting the result into Eq.~(\ref{AP2.Id+Ir})
  and rearranging the terms by their $\varphi_*$-dependence 
  into the form of Eq.~(\ref{ChiR//2phi*}), 
  we obtain the longitudinal and transverse {DC}s
  as functions of $\vartheta$.
Using $f_\pm$ to express these correlations as \begin{equation}
  \Xl  = {\frac12}\left({f_+} + {f_-}\right), \quad 
  \Xtr = {\frac12}\left({f_+} - {f_-}\right)  \notag 
\end{equation}
  in accordance with Eq.~(\ref{f//RR}), 
  we have 
\begin{align}
  f_+ &= 
  \frac{S}{2\pi} \ell_0^2 \left[
  E_1(\vartheta^2) + \frac{E_1({\vartheta^2}/\muR)}{\muR} 
  \right]
  \label{f+//th2}, 
  \\
  f_- &= 
  \frac{S}{2\pi} \ell_0^2 \times 
  \frac{e^{-\vartheta^2} - e^{-{\vartheta^2}/\muR}}{\vartheta^2}
  \label{f-//th2},  
\end{align}%
  where $E_1(~\cdot~)$ 
  denotes the exponential integral
  defined by \cite{Arfken.Book2013}
\begin{equation}
  E_1(w) 
  = \int_w^\infty \frac{\exp(-z)}{z}\D{z}  %
  \label{expInt}.
\end{equation}
  
With Eq.~(\ref{f+//th2}) in hand, 
  we can show readily 
  that $f_+$ behaves logarithmically on shorter length\-scales.
In terms of $\vartheta$ 
  defined in Eq.~(\ref{sim.theta}), 
  by the ``shorter length\-scales'' we mean 
  the range of $\di[\relax] = \ell_0\xi_*$ 
  satisfying 
  both $\vartheta \ll 1$ and 
  $\di[\relax]> \ell_0$.
Using 
\begin{equation}
  E_1(w) \simeq -\ln{w} -\gmEM + w - \frac{w^2}{4} + \cdots
\end{equation}   
  for small $w$ (with $\gmEM \approx 0.5772$ denoting 
  the Euler--Mascheroni constant), 
  it is straightforward to obtain 
\begin{equation}
  \frac{f_+}{(S/2\pi)\ell_0^2}
  = -\frac{2(1+\muR)}{\muR}\ln\vartheta 
  + \Order(1)
  \label{f+//th.small}
\end{equation}
  for small $\vartheta$.
It is also easy 
  to show 
\begin{equation}
  \frac{f_-}{(S/2\pi)\ell_0^2}
  = \frac{1-\muR}{\muR} + \Order(\vartheta^2) 
  \label{f-//th.small}
\end{equation}
  from Eq.~(\ref{f-//th2}).
The presence of the logarithmic behavior 
  in Eq.~(\ref{f+//th.small})
  and its absence from Eq.~(\ref{f-//th.small})
  are qualitatively consistent with the numerical results
  in Fig.~\ref{Fig:f//th} 
  for small $\vartheta\;(< 0.1)$.
Besides,
  with regard to the behavior of $f_-$ for large $\vartheta$,
  we find Eq.~(\ref{f-//th2}) 
  to be consistent with our numererical results, as shown in Fig.~\ref{Fig:th2*f-}.

The analytical expressions of $f_\pm$ 
  in Eqs.~(\ref{f+//th.small}) and (\ref{f-//th.small})
  allow us to extract information of elasticity 
  from simulational and experimental data 
  of the particle system, 
  on the assumption that the elastic model is quantitatively valid. 
They may provide useful alternatives to Eq.~(\ref{muR=Cd/Cr}),  
  as $\ChiR(\di,\tD)$ is easier to compute 
  than its Fourier counterpart used in Eq.~(\ref{muR=Cd/Cr}).


Starting from the same elastic modeling
  and following basically the same procedure, 
  we can also calculate the SC. 
The only difference 
  is that the Alexander--Pincus formula (\ref{eqs:I.AP2}) 
  is replaced with Eq.~(\ref{gm//C.H}).
In polar coordinates, 
  the calculation of SC 
  reduces to $I_{0,2}$ and $I_{4,2}$ 
  in Appendix~\ref{app:intK.Bessel.gauss};
  the result reads
\begin{multline}
  \frac{\chi_\gm^{}}{S/(\pi\xi_*^2)}
  = 
  - \left( \vartheta^2 e^{-\vartheta^2} 
    + \frac{\vartheta^2 e^{-\vartheta^2/\muR}}{\muR^2}
  \right)
  \brX 
  + \left[ 
  Q_\gm(\vartheta^2)     - \dfrac{Q_\gm(\vartheta^2/\muR)}{\muR}
  \right]\cos{4\varphi_*}%
  \label{chiGm//th},\hFil 
\end{multline}
  where we have defined $Q_\gm(w) = (6w^{-1}+4+w)e^{-w}$.
This analytical expression 
  is plotted in Fig.~\ref{Fig:chiGm} as a 2D color map,
  and delineated in Fig.~\ref{Fig:cmp3}
  as a function of $\di[\relax] \;({}= 2\sqrt{\Dc\tD}\,\vartheta)$
  for $\varphi _* = 0$ and $\varphi _* = \pi /4$.

Obviously, 
  Eq.~(\ref{chiGm//th}) 
  includes $\cos{4\varphi_*}$ in the form of Eq.~(\ref{gm.0+4}).
This $\varphi_*$-dependence 
  leads to the well-known cross-shaped ``flower'' pattern 
  shown in Fig.~\ref{Fig:chiGm}.

The $\xi_*$-dependence of Eq.~(\ref{chiGm//th}) 
  is less obvious.
Recalling Eq.~(\ref{gm.0+4}) 
  to denote the two angular modes 
  with $\chi_\gm^{(0)}$ and $\chi_\gm^{(4)}$, 
  we expand them for small $\vartheta$ 
  to find 
\begin{align}
  \chi_\gm^{(0)} &= \frac{S}{\pi\xi_*^2}\left[ 
  -\frac{1+\muR^2}{\muR^2} \vartheta^2 + \Order(\vartheta^4)
  \right]
  = \Order\left( ({\Dc_*}t)^{-1} \right)  
  \label{fm0//th.small}, 
  \\
  \chi_\gm^{(4)} &= \frac{S}{\pi\xi_*^2}\left[ 
  \frac{2(1-\muR)}{\muR}+\Order(\vartheta^6)
  \right]
  \simeq \frac{2(1-\muR)S}{\pi\muR\xi_*^2}
  \label{fm4//th.small}; 
\end{align}
  here it should be noted that, 
  since $\di[\relax] > \ell_0$, 
  the smallness of $\vartheta$ implies largeness of ${\Dc_*}t$. 

It is easy to verify 
  that the above expressions for DCs and SCs 
  are consistently interrelated by Eqs.~(\ref{eqs:L}).
In particular, 
  Eq.~(\ref{f+//th.small}) implies 
  that $f_+$ is asymptotically a linear combination 
  of $\{{\ln\xi_*}, 1\}$, 
  which belongs to the null space of $\hat{L}_+$ 
  and therefore most of the information of $f_+$ 
  is lost from $\chi_\gm^{(0)}$.
Contrastively, the information of $f_-$ in Eq.~(\ref{f-//th.small}) 
  is transmitted by $\hat{L}_-$ to $\chi_\gm^{(4)}$ properly, 
  in the sense 
  that both $f_-$ in Eq.~(\ref{f-//th.small}) 
  and $\chi_\gm^{(4)}$ in Eq.~(\ref{fm4//th.small})
  contain the same factor $(1-\muR)/\muR$.


\subsection{Logarithmic regime as a reflection of cage effect}
\label{subsec:log.cage}

We have seen 
  that the logarithmic behavior of the DC tensor 
  can be derived from the elastic modeling. 
Here the elastic modeling 
  means assumption of approximate dynamics 
  leading to Eqs.~(\ref{eqs:Corr.Delta.exp}) 
  for $C_a^\Delta$ (with $a\in\{\mathrm{d},\mathrm{r}\}$)
  and allowing derivation of Eq.~(\ref{f+//th.small}), 
  which implies that the $\xi_*$-dependence of $f_+$
  is asymptotically logarithmic
  for distances shorter than $2\sqrt{{\Dc_*}t}$.

Now we will extend this result to a wider class of modeling,
  having various kinds of viscoelastic or elasto\-plastic models 
  for $\psiD$ and $\psiR$ in view.
Instead of analyzing some specific model in detail, 
  we assume only that $C_a^\Delta$ derived from the modeling  
  satisfies certain conditions specified below.
On this assumption, 
  we will show $f_+$ to behave asymptotically as 
\begin{equation}
  \frac{f_+}{\sigma^2} 
  = -{C_+}\ln\frac{\xi_*}{\ld}  %
  \label{f+//C+}
\end{equation}
  for $\xi_* \gg \ld$, 
  where $C_+$ and $\ld$ are constant with regard to $\xi_*$ 
  but possibly dependent on $\tD$. 
The result of the elastic modeling in Eq.~(\ref{f+//th.small})
  is a special case of Eq.~(\ref{f+//C+}). 

To proceed, 
  we need some minimal prescription for
  $C_a^\Delta = C_a^\Delta(k,\tD)$.
As is evident from its definition in Eq.~(\ref{C.Delta}), 
  the ``initial value'' of $C_a^\Delta$ for $\tD=0$ is zero, 
  and it is natural to expect 
  that $C_a^\Delta(k,\tD)$ is a growing function of $\tD$.
Taking this $\tD$-dependence into account, 
  here we consider the following two types of the $k$-dependence:
\begin{enumerate}
\item For shorter wavelengths, 
      $C_a^\Delta(k,\tD)$ grows with lapse of $\tD$ 
      and reaches a constant saturation value independent of $k$.
      On the other hand, 
      $C_a^\Delta(k,\tD)$ vanishes 
      in the limit of long waves ($k\to0$).
\item Behaving in proportion to $k^2$ 
      over the entire range of wave\-numbers, 
      $C_a^\Delta(k,\tD)$ diverges 
      for shorter waves and for $\tD\to\infty$.
\end{enumerate}
These are two typical behaviors of $\Cd^\Delta$ and $\Cr^\Delta$ 
  that we have encountered in Ref.~\cite{Ooshida.PRE94}.
The first type leads to Eq.~(\ref{f+//C+}), 
  while the second type does not.
Note that the classification is not comprehensive, 
  but discussion on other types of behavior 
  is out of the scope of the present work. 

\subsubsection{First type: caged behavior with a saturation value}

The logarithmic behavior in Eq.~(\ref{f+//C+}) 
  is shown to result from the first type mentioned above, 
  characterized by the emergence of a constant saturation value. 
From the viewpoint of particle dynamics, 
  the saturation of $C_a^\Delta(k,\tD)$ 
  is understood as reflection of the cage effect 
  on the rotational or dilatational modes 
  of meso\-scopic deformation.
Note that $\tauA$, 
  usually regarded 
  as representing the lifetime of the cage,
  is measured with $\Fs(k,\tD)$ which is a single-particle quantity;
  this means that $\tauA$ measures 
  the collapse of the cage on the length\-scale of $\sigma$, 
  but does not eliminate the possibility 
  that, for other length\-scales,
  the cage effect may last for times longer than $\tauA$.

The assumptions of the saturation stated above  
  can be reformulated more precisely 
  by postulating the existence 
  of a (nondimensionalized) length\-scale, $\ld = \ld(\tD)$, 
  such that 
\begin{equation}
  C_a^\Delta(k,\tD) \sim 
  \begin{cases}
  D_* k^2 \tD    & {(0 < k \ll 1/\ld)} \\
  {{S_a^\sharp}}  & {({1/\ld} \ll k < \kmax)} 
  \end{cases}
  \label{ve.caged}
\end{equation}  
  where ${S_a^\sharp}$ is the saturation value 
  (independent of $\tD$ and $k$),
  and $\kmax$ is the cutoff wavenumber 
  corresponding to the length\-scale ${}\sim {\ell_0}$.
For the sake of simplicity, however, 
  we treat $\kmax$ as if it is infinitely large.   
It is also assumed that $0 \le {C_a^\Delta(k,\tD)}\le{S_a^\sharp}$
  over the entire range of $k$. 
We note 
  that Eqs.~(\ref{eqs:Corr.Delta.exp}), given by the elastic modeling, 
  satisfy the assumptions in Eq.~(\ref{ve.caged}), 
  with $\ld \sim 2\sqrt{\Dc_*\tD}$,   
  ${{\Sd^\sharp}} = S$ and ${{\Sr^\sharp}} = S/\muR$, 
  and therefore the behavior of $\Cd^\Delta$ and $\Cr^\Delta$ 
  given by the elastic modeling 
  belongs to this type.   

Unlike Eqs.~(\ref{eqs:Corr.Delta.exp}),
  $C_a^\Delta(k,\tD)$ in general 
  is not necessarily reducible to the form 
  tractable with the integrals 
  given in Appendix~\ref{app:intK.Bessel.gauss}.
Therefore we must go back to the Alexander--Pincus formula, 
  given as Eqs.~(\ref{eqs:AP2.polar}) in polar coordinates.
To denote the integrals contained in Eqs.~(\ref{eqs:AP2.polar})
  conveniently, 
  we define 
\begin{equation}
  I_a^{(m)}(\xi_*) 
  = \int_0^\infty C_a^\Delta(k,\tD) J_m(k\xi_*) \frac{\D{k}}{k}  
  \label{Ia.m}
\end{equation}
  where $a\in\{\mathrm{d},\mathrm{r}\}$
  and $m$ is an integer; 
  for evaluation of Eqs.~(\ref{eqs:AP2.polar})
  we need only $I_a^{(0)}$ and $I_a^{(2)}$.

To demonstrate the presence of the logarithmic regime, 
  let us evaluate the integral in Eq.~(\ref{Ia.m}) 
  for the shorter-scale range of $\xi_*$
  specified as $ \kmax^{-1} \ll \xi_* \ll \ld $.
In consideration of the factor $1/k$ in the integrand, 
  we split the integral at the wavenumber $1/\ld$, 
  as 
\begin{multline}
  I_a^{(m)}(\xi_*) 
  = \int_0^{1/\ld}      C_a^\Delta(k,\tD) J_m(k\xi_*) \frac{\D{k}}{k}\brX   
  + \int_{1/\ld}^\infty C_a^\Delta(k,\tD) J_m(k\xi_*) \frac{\D{k}}{k}  
  \label{Ia.m.split}.\hFil 
\end{multline}  
For $m\ge1$, 
  it is easy to show 
  that the contribution from the vicinity of $k=0$ is negligible,
  because $J_m(k\xi_*)$ is small in proportion to $(k\xi_*)^m$ 
  for small $k$.
The second integral can be evaluated 
  by changing the variable of integration from $k$ to $q = k\xi_*$
  and noticing that the lower bound of $q$ equals $\xi_*/\ld \ll 1$.
In particular, 
  for $m=2$ we have
\begin{equation}
  I_a^{(2)}(\xi_*) 
  \simeq 
  {S_a^\sharp} \int_{1/\ld}^\infty J_2(k\xi_*) \frac{\D{k}}{k}
  = {\frac12} {S_a^\sharp} 
  \label{Ia2//S}
\end{equation}
  with the terms of $\Order\left({(\xi_*/\ld)^2}\right)$ 
  discarded.

The case of $m=0$ 
  requires a more careful treatment. 
The first integral on the right-hand side of Eq.~(\ref{Ia.m.split}) 
  does not vanish; 
  rather, it is shown to remain finite for small $\xi_*$.
Taking the $k$-dependence of $C_a^\Delta(k,\tD)$  
  prescribed in Eq.~(\ref{ve.caged}) into account, 
  at the lowest order of the long\-wave approximation, 
  we can evaluate the integral as 
\begin{multline}
  \int_0^{1/\ld} C_a^\Delta(k,\tD) J_0(k\xi_*) \frac{\D{k}}{k}
  \simeq \int_0^{1/\ld} \frac{C_a^\Delta(k,\tD)}{k} \D{k} \brX 
  \simeq {D_*\tD} \int_0^{1/\ld} k\,\D{k}
  = \frac{D_*\tD}{2\left\{\ld(\tD)\right\}^2} 
  \label{Ia0.LW}
\end{multline} 
  to find it finite.
Consideration of terms in higher order of $k$  
  in the long\-wave approximation
  does not change the conclusion 
  that the integral gives a finite value.

To evaluate the second integral, 
  we make use of $(\ln{k\ld})' = 1/k$, 
  with the prime denoting $\partial/\partial{k}$. 
Upon integration by parts,  
  we have 
\begin{align}
  &\int_{1/\ld}^\infty C_a^\Delta(k,\tD) J_0(k\xi_*) \frac{\D{k}}{k}
  \notag \\
  &= 
  \int_{1/\ld}^\infty C_a^\Delta(k,\tD) 
  J_0(k\xi_*) (\ln{k\ld})  ' \D{k}
  \notag \\ 
  &\simeq {-{S_a^\sharp}} 
  \int_{1/\ld}^\infty [ J_0(k\xi_*) ]' (\ln{k\ld}) \D{k}
  \label{Ia0.logK}. 
\end{align}
Subsequently, 
  changing the variable from $k$ to $q = k\xi_*$,  
  we find 
\begin{gather*}
  [ J_0(k\xi_*) ]' \D{k} = \DD{[J_0(q)]}{q} \D{q} = -J_1(q)\D{q}
  \relax, \\ 
  \ln{k\ld} = \ln\frac{q\ld}{\xi_*} 
  = \ln{q} - \ln{\frac{\xi_*}{\ld}}
\end{gather*}
  so that 
\begin{align}
  \int_{1/\ld}^\infty [ J_0(k\xi_*)     ]  ' (\ln{k\ld}) \D{k}
  &= 
  -\int_{\xi_*/\ld}^\infty
  \left( \ln{q} - \ln{\frac{\xi_*}{\ld}} \right)
  J_1(q) \D{q} 
  \notag \\ 
  &\simeq
  \int_0^\infty
  \left( \ln{\frac{\xi_*}{\ld}} - \ln{q} \right)
  J_1(q) \D{q} 
  \label{Ia0.logXi}
\end{align}
  for $\xi_* \ll \ld$.
Evaluating the integral 
  and combining the result with Eq.~(\ref{Ia0.LW}), 
  finally we obtain 
\begin{equation}
  I_a^{(0)}(\xi_*) 
  \simeq 
  - {S_a^\sharp}\ln\frac{\xi_*}{\ld} + \Order(1)
  \label{Ia0//S*log}.
\end{equation}
Note that the term of $\Order(1)$ in Eq.~(\ref{Ia0//S*log}) 
  includes the contribution of Eq.~(\ref{Ia0.LW})
  which is constant with regard to $\xi_*$ 
  but may or may not depend on $\tD$.
Without loss of generality,
  the term of $\Order(1)$ can be set equal to zero, 
  because Eq.~(\ref{ve.caged}) leaves room 
  for redefining $\ld$ to absorb it.


Now suppose 
  that a certain visco\-elastic modeling under consideration 
  gives caged dynamics, both for $\Cd^\Delta$ and for $\Cr^\Delta$, 
  so that Eqs.~(\ref{Ia2//S}) and (\ref{Ia0//S*log}) 
  are valid for both modes.
Using Eqs.~(\ref{Ia2//S}) and (\ref{Ia0//S*log}), 
  we can evaluate the DC tensor 
  given by Eqs.~(\ref{eqs:AP2.polar}),
  from which we can extract $f_\pm$ 
  by Eqs.~(\ref{ChiR//2phi*}) and (\ref{f//RR}).
The result reads
\begin{subequations}%
\begin{align}
  f_+ &\simeq 
  -\frac{\ell_0^2}{\pi} \left( {\Sd^\sharp} + {\Sr^\sharp} \right)
  \ln\frac{\xi_*}{\ld} %
  \label{f+//Sd+Sr},
  \\
  f_- &\simeq
  \frac{\ell_0^2}{2\pi} \left( -{\Sd^\sharp} + {\Sr^\sharp} \right)
  \label{f-//S}.
\end{align}
\label{eqs:f//caged}%
\end{subequations}%
Thus $f_+$ is found to behave logarithmically
  as in Eq.~(\ref{f+//C+}), 
  with $C_+ = ({\ell_0^2}/(\pi\sigma^2))({\Sd^\sharp} + {\Sr^\sharp})$.

As was noted earlier in connection with Eq.~(\ref{ve.caged}), 
  the elastic modeling in the previous subsection 
  is a special case of caged dynamics
  satisfying Eq.~(\ref{ve.caged}).
It is easy to confirm 
  that setting ${{\Sd^\sharp}} = S$ and ${{\Sr^\sharp}} = S/\muR$
  in Eqs.~(\ref{f+//Sd+Sr}) and (\ref{f-//S}),
  with $\ld \sim 2\sqrt{{\Dc_*}\tD}$,   
  reproduces Eqs.~(\ref{f+//th.small}) and (\ref{f-//th.small}).

\subsubsection{Second type: uncaged behavior for shorter waves}

As a contrast to the caged behavior 
  characterized by Eq.~(\ref{ve.caged}), 
  here we consider another type of $k$-dependence, 
  in which $C_a^\Delta(k,\tD)$ diverges  
  without being saturated by cage effect.
More specifically, 
  we assume 
\begin{equation}  
  C_a^\Delta(k,\tD) \simeq  D_* k^2 \tD 
  \label{ve.uncaged}
\end{equation}  
  for the entire range of $k$,
  with the case of $a = \mathrm{r}$ in mind.
This type of behavior 
  is seen in $\Cr^\Delta$ 
  when the fluctuating dynamics of $\psiR$ 
  are equivalent to those of free Brownian motion 
  without restoring force, 
  as is the case in the linear equation of $\psiR$
  studied in Subsec.~IV-C of Ref.~\cite{Ooshida.PRE94}.
  
The integrals needed for evaluation of the DC tensor 
  are still given by Eq.~(\ref{Ia.m}) with $m=2$ and $m=4$.
For $C_a^\Delta$ now we use Eq.~(\ref{ve.uncaged}),  
  which implies 
  oscillatory divergence of the integrand for $k\to\infty$.
Handling the divergence 
  with the standard technique of convergence factor, 
  we find that $I_a^{(0)}(\xi_*)$ vanishes for $\xi_* > 0$
  (the integral reduces to the delta function of $\xi_*$).  
For $m=2$,
  using the convergence factor $e^{-\epsilon{k}}$ 
  and taking the limit of $\epsilon\to+0$,
  we obtain 
\begin{multline}
  \int_0^\infty 
  \Cr^\Delta(k,t-s) J_2(k\xi_*) e^{-\epsilon{k}}\frac{\D{k}}{k}
  \brX 
  =
  D_*\,(t-s) \int_0^\infty J_2(k\xi_*) e^{-\epsilon{k}} k \D{k}
  \to 
  \frac{2 D_*\,(t-s)}{\xi_*^2}
  \label{int2r.k2},\hFil  
\end{multline}
  which diverges for $\xi_*\to 0$.

The results obtained above 
  can be applied to the case of linearized dynamics 
  studied in Subsec.~IV-C of Ref.~\cite{Ooshida.PRE94}.
In this case, 
  the rotational mode is uncaged 
  so that $\Cr^\Delta$ behaves according to Eq.~(\ref{ve.uncaged}), 
  while $\Cd^\Delta$ is subject to saturation 
  as in Eq.~(\ref{ve.caged}).
Taking $I_{\mathrm{r}}^{(0)}(\xi_*) = 0$ into account, 
  we find 
\begin{subequations}%
\begin{align}
  f_+ &\simeq
  -\frac{\ell_0^2}{\pi} {\Sd^\sharp} \ln\frac{\xi_*}{\ld} %
  \label{f+//S},
  \\
  f_- &\simeq
  \frac{2{\ell_0^2} D_*\tD}{\pi{\xi_*^2}}
  \label{f+//inv2}
\end{align}%
\label{eqs:f//uncaged}%
\end{subequations}%
  in this case.
The result 
  is consistent with Eq.~(4.37) of Ref.~\cite{Ooshida.PRE94}.

\subsection{Quantitative data analysis of the logarithmic regime}
\label{subsec:log.analysis}

With Eqs.~(\ref{eqs:f//caged}) and (\ref{eqs:f//uncaged})
  relating $f_\pm$ to the two types of behavior of $C_a^\Delta$, 
  now let us analyze the numerical data of $f_\pm$. 
The analysis may allow us 
  to extract information of the assumed dynamics, 
  such as the elastic moduli.

\begin{figure}
\centering
\raisebox{5.0cm}{(a)}\includegraphics[clip,width=7.0cm]{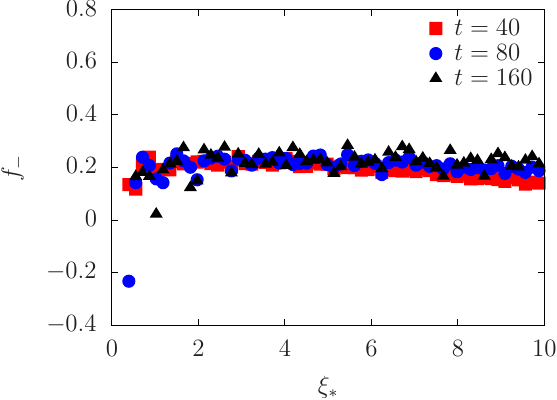}\quad
\raisebox{5.0cm}{(b)}\includegraphics[clip,width=7.0cm]{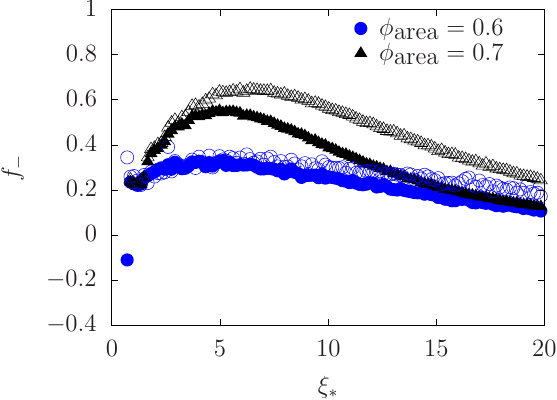}
\caption{\label{Fig:f-}%
    $f_-$ nondimensionalized with $\sigma^2$
  and plotted against $\xi_* = \di[\relax]/\ell_0$.
  (a) Plots in the case of $\phiA = 0.5$
  for three different values of the time interval:
  $t = 40$, $80$ and $160$.
  (b) Plots in the cases of $\phiA = 0.6$ and $\phiA = 0.7$,
  for $t = 80$ (closed symbols) and $t = 160$ (open symbols).
}
\end{figure}

The main question here
  is whether $\Cr^\Delta$ is caged or uncaged.
This is answered 
  by checking which of Eq.~(\ref{f-//S}) and Eq.~(\ref{f+//inv2}) 
  is closer to the actual behavior of $f_-$.
From Fig.~\ref{Fig:f-}
  where $f_-$ is plotted against $\xi_* = \di[\relax]/\ell_0$, 
  we can judge that $\Cr^\Delta$ is caged 
  for all the three values of $\phiA$ studied here. 
As is shown in Fig.~\ref{Fig:f-}(a), 
  the values of $f_-$ in the case of $\phiA = 0.50$  
  is nearly constant (i.e.\ 
  independent of $\xi_*$ and $\tD$), 
  except for deviation at very small distance.
We see from Fig.~\ref{Fig:f-}(b) 
  that $f_-$ for $\phiA = 0.60$ is also nearly constant.
The constancy of $f_-$ 
  seems to be consistent with Eq.~(\ref{f-//S}).
The case of $\phiA = 0.70$
  is difficult to interpret, 
  as $f_-$ is nor constant nor proportional to $\xi_*^{-2}$,
  but at least we can eliminate 
  the divergent behavior in Eq.~(\ref{f+//inv2}) 
  predicted for the case of uncaged $\Cr^\Delta$.
Thus we can regard Eqs.~(\ref{eqs:f//caged}) 
  as a reasonable approximation of $f_\pm$, 
  which provides supporting evidence 
  that the cage effect may last longer than $\tauA$ 
  at some length\-scales greater than $\ell_0$.

As a corollary of the above result, 
  we find that $f_+$ dominates over $f_-$ for $\xi_* \ll \ld$.
On the basis of this dominance,
  we can estimate the magnitude of the DC 
  at the typical inter\-particulate distance $\ell_0$, 
  by extrapolating 
  the asymptotic behavior of $f_+$ in Eq.~(\ref{f+//Sd+Sr})
  to $\xi_* = 1$.
This is approximately the value of MSD 
  that the particle would have 
  if it were eternally confined 
  in the cage of the neighboring particles.
Denoting it with $\Av{\mb{R}^2}_{\text{caged}}$,  
  for $\ld = \ld(t) \gg 1$ 
  we have  
\begin{equation}
  \Av{\mb{R}^2}_{\text{caged}}
  \approx 
  \Ev{f_+}{{\xi_*}=1} 
  \simeq 
  \frac{\ell_0^2}{\pi} 
  \left( {\Sd^\sharp} + {\Sr^\sharp} \right) \ln{\ld(t)} %
  \label{MSD.caged}, 
\end{equation}
  which gives the time-dependent version 
  of the Mermin--Wagner fluctuation, 
\begin{equation}
  \Av{\mb{R}^2}_{\text{caged}} \propto 
  \frac{\ell_0^2}{2\pi} 
  \left( {\Sd^\sharp} + {\Sr^\sharp} \right) \ln({D_*}\tD)
  \label{MSD//log.tD},
\end{equation}
  if $\ld(t)$ behaves as $\sqrt{{D_*}\tD\mathstrut}$ 
  and is greater than unity but smaller than the system size.

\begin{figure}
  \includegraphics[clip,width=7.0cm]{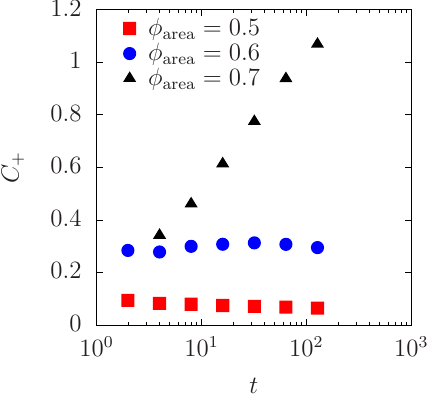}%
  \caption{\label{Fig:C+}%
    ${C_+ = -\D({f_+/\sigma^2})/\D(\ln\xi_*)}$ 
    for $\phiA = 0.5$, $0.6$ and $0.7$, plotted against $t$. 
  }
\end{figure}
\begin{figure}%
  \centering
  \includegraphics[clip,width=7.0cm]{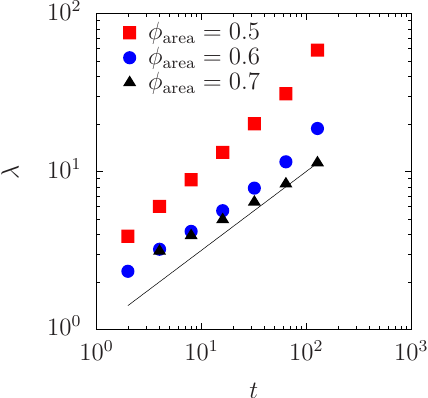}%
  \caption{\label{Fig:ld//t}%
    Length scale $\ld$ 
    evaluated by fitting Eq.~(\ref{f+//C+}) 
    to the numerical data of $f_+ = \Xl+\Xtr$. 
    The solid line indicates $\ld \propto \tD^{1/2}$.
  }%
\end{figure}%

\begin{figure}
  \centering
  \includegraphics[clip,width=7.0cm]{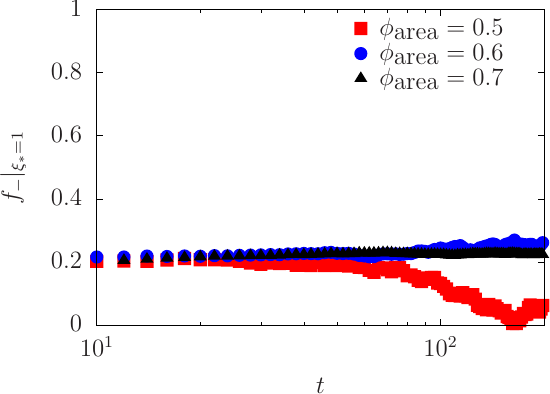}
  \caption{\label{Fig:C-}%
    The values of 
    ${f_- \simeq {\frac12}{\sigma^2}{C_-}}$ at $\xi_* = 1$ 
    for $\phiA = 0.5$, $0.6$ and $0.7$, 
    plotted against $t$\protect,
    from which $C_-$ is estimated.
  }
\end{figure}

Next, we discuss 
  how to extract quantitative information from the data of $f_+$
  by means of Eq.~(\ref{f+//Sd+Sr}). 
More specifically,
  we aim to evaluate ${\Sr^\sharp}$,
  which is expected to carry information of the shear modulus.
Noticing 
  that Eq.~(\ref{f+//Sd+Sr}) is in the form of Eq.~(\ref{f+//C+}), 
  we fit this form 
  to the data of $f_+$ as a function of $\xi_*$,   
  within the range 
  in which the logarithmic behavior is observed, 
  to obtain 
  $C_+ = -\D(f_+/\sigma^2)/\D(\ln\xi_*)$ and $\ld$ 
  as fitting parameters.
In other words,  
  we regard Eq.~(\ref{f+//C+}) 
  as the definition of $C_+$ and $\ld$ through fitting. 
The values of $C_+$ and $\ld$ 
  thus obtained for various time intervals, $\tD$, 
  are plotted in Fig.~\ref{Fig:C+} and Fig.~\ref{Fig:ld//t}, 
  respectively. 

Let us examine 
  the values of $C_+$ in Fig.~\ref{Fig:C+}.
For $\phiA = 0.5$ and $\phiA = 0.6$, 
  $C_+$ is nearly independent of $\tD$.
In the case of $\phiA = 0.7$,  
  $C_+$ grows as $\tD$ increases; 
  the limitation due to the finite system size 
  makes it difficult to determine 
  whether $C_+$ converges to some limiting value 
  or grows unlimitedly. 
If it actually converges, 
  we may say that the limiting value of $C_+$ 
  increases as $\phiA$ is increased. 
 
From the logarithmic plot in Fig.~\ref{Fig:ld//t}, 
  we can read that the length scale $\ld = \ld(\tD)$ grows 
  nearly in proportion to $\sqrt{\tD}$.
This diffusive behavior of $\ld(\tD)$
  is consistent with the prediction of the elastic model 
  that the DCs are expressed 
  in terms of the similarity variable $\vartheta$ 
  in Eq.~(\ref{sim.theta}).
It also supports one of the assumptions 
  underlying Eq.~(\ref{MSD//log.tD})
  about $\Av{\mb{R}^2}_{\text{caged}}$.
In regard to the $\phiA$-dependence of $\ld$, 
  we find that $f_+$ becomes less diffusive 
  as $\phiA$ is increased.

Lastly, we ask 
  whether quantitative information  
  extracted from the data is consistent with the elastic modeling.
Such information must be contained, at least partially, 
  in the values of $C_+$ obtained from Fig.~\ref{Fig:C+}.
Since $C_+$ equals
  $(\ell_0^2/(\pi\sigma^2))({\Sd^\sharp} + {\Sr^\sharp})$
  according to Eq.~(\ref{f+//Sd+Sr}),
  it should be possible 
  to obtain therefrom the value of ${\Sr^\sharp}$, 
  which is related to the shear modulus, 
  if ${\Sd^\sharp}$ is known somehow.
There are at least two possible ways to evaluate ${\Sd^\sharp}$: 
  we may substitute for it 
  the static structure factor of the density, $S$,  
  or we may use Eq.~(\ref{f-//S}) 
  to estimate $-{\Sd^\sharp} + {\Sr^\sharp}$ on the basis of $f_-$.

Let us discuss 
  the first choice (i.e.~evaluation from $S$ and $C_+$).
We can estimate 
\begin{equation}
  {\Sr^\sharp} 
  \approx \frac{\pi\sigma^2}{\ell_0^2} {C_+} - S 
  = 4\phiA {C_+} - S 
  \label{Sr//S0}
\end{equation}  
  from the value of $S$ in Table~\ref{Tab:S}  
  and $C_+$ read from Fig.~\ref{Fig:C+}.
For $\phiA = 0.50$, 
  we find ${\Sr^\sharp}$ to be vanishing small; 
  although literal interpretation of Eq.~(\ref{Sr//S0}) 
  gives ${\Sr^\sharp} = -0.02$, 
  we must remember that ${\Sr^\sharp}$ cannot be negative.
In the case of $\phiA = 0.60$ 
  we have ${\Sr^\sharp} \approx 0.6$ according to Eq.~(\ref{Sr//S0}), 
  and ${\Sr^\sharp} \approx 2.8$ 
  if we take $C_+ \approx 1$ for $\phiA = 0.70$.

The second choice 
  turns out to be perplexing.
In view of the approximate constancy of $f_-$ 
  within a certain range of $\xi_*$ shown in Fig.~\ref{Fig:f-}, 
  we write the constant
  as $f_- \simeq ({\sigma^2}/2) C_-$ 
  in parallel with $C_+$.
The values of $C_-$, shown in Fig.~\ref{Fig:C-},
  are nearly independent of $\tD$ %
  (except for the decay after $t=70$ in the case of $\phiA = 0.5$).
In terms of $C_-$ thus defined,
  Eq.~(\ref{f-//S}) reads 
  $C_- \simeq ({\ell_0^2}/(\pi\sigma^2)) (-{\Sd^\sharp} + {\Sr^\sharp})$ 
  , 
  so that Eqs.~(\ref{eqs:f//caged}) yield
\begin{subequations}
\begin{align}
  {\Sd^\sharp} 
  &\approx 
  \frac{\pi\sigma^2}{2\ell_0^2} \left( {C_+} - {C_-} \right)
  = 2\phiA \left( {C_+} - {C_-} \right)
  \label{Sd//C+-},
  \\
  {\Sr^\sharp} 
  &\approx 
  \frac{\pi\sigma^2}{2\ell_0^2} \left( {C_+} + {C_-} \right)
  = 2\phiA \left( {C_+} + {C_-} \right)
  \label{Sr//C+-}.
\end{align}%
\label{eqs:Sa//C+-}%
\end{subequations}%
Since both ${\Sd^\sharp}$ and ${\Sr^\sharp}$ must be positive,
  we should have $0 < C_- < C_+$, 
  but the values of $C_-$ read from Fig.~\ref{Fig:C-}
  ($C_- \simeq 2 f_-/\sigma^2 \approx 0.4$
  both for $\phiA = 0.50$ and $\phiA = 0.60$)
  contradict this inequality,
  as the value is greater than $C_+$ in Fig.~\ref{Fig:C+}
  ($C_+ < 0.1$ for $\phiA = 0.5$ and $C_+ \approx 0.3$ for $\phiA = 0.6$)%
  .
In the case of $\phiA = 0.70$, 
  the ``second choice'' estimation 
  from $C_\pm$ by way of Eq.~(\ref{Sr//C+-})
  gives ${\Sr^\sharp} \approx 2.0$ 
  (if we take $C_+ \approx 1$ and $C_- \approx 0.4$), 
  which is somewhat smaller than ${\Sr^\sharp} \approx 2.6$ 
  obtained from $S$ and $C_+$ (the first choice).


In spite of the quantitative inconsistency, 
  however, 
  the two ways of estimating ${\Sr^\sharp}$ as a function of $\phiA$
  are in agreement about qualitative tendency.
As $\phiA$ is increased from $0.6$ to $0.7$, 
  the value of ${\Sr^\sharp}$ estimated by Eq.~(\ref{Sr//S0})
  increases from about $0.6$ to $2.6$, 
  and also the estimation by Eq.~(\ref{Sr//C+-}) 
  shows an increase from $0.9$ to $2.0$.
This tendency 
  suggests the need of something more sophisticated 
  than the simple elastic modeling, 
  because, according to Eq.~(\ref{Cr.Delta.exp})
  derived from Eq.~(\ref{dt.psiR.muR}), 
  ${\Sr^\sharp}$ is inversely proportional to the shear modulus
  and therefore supposed to be a \emph{decreasing} function of $\phiA$.
The present analysis, 
  revealing that ${\Sr^\sharp}$ is an \emph{increasing} function of $\phiA$
  for $\phiA \le 0.7$ and $\tD \gg \tauA$,  
  provides useful information 
  for searching suitable models 
  about the dynamics of $\Cd^\Delta$ and $\Cr^\Delta$.


\section{Concluding remarks}
\label{sec:conc}

We have investigated 
  three questions about space--time correlations 
  in a model colloidal liquid.
First, we asked 
  whether the displacements have nontrivial correlations 
  even in liquids that are only slightly glassy.
The answer is affirmative: 
  this is demonstrated 
  by computing $\ChiR$, the DC tensor defined in Eq.~(\ref{ChiR}),
  and $\chi_\gm$, the SC defined by Eq.~(\ref{chiGm=}), 
  from simulation data of a two-dimensional model colloidal liquid 
  with $\phiA = 0.50$, $0.60$ and $0.70$.
This is consistent 
  with the experimental observation 
  by Illing \textit{et al.}~\cite{Illing.PRL117}.
The time dependence of these correlations 
  is not linear with regard to the time interval $\tD$,  
  as would be expected
  if the detected correlations were explained 
  as accumulation of many events,  
  but rather described in terms of a similarity variable 
  $\vartheta$ in Eq.~(\ref{sim.theta}), 
  indicating 
  the presence of the diffusive correlation length. 

Secondly,
  we asked whether DC and SC are equivalent.
The answer is negative:
  to answer this question,
  we have derived Eqs.~(\ref{eqs:L}) 
  as a relation between the two kinds of correlations,
  by treating the displacement field 
  with the label variable formulation.
The computed values of DC and SC 
  are then shown to be consistent 
  with Eqs.~(\ref{eqs:L}).

The relation in Eqs.~(\ref{eqs:L}) 
  takes the form of a linear mapping 
  from the DC to the SC, 
  which is expressed 
  by means of two linear operators, $\hat{L}_\pm$, 
  of Euler--Cauchy type.
This mapping is non-invertible: 
  we can discover 
  $\chi_\gm$  %
  from the components of the DC,
  $f_\pm = \Xl \pm \Xtr$, 
  but cannot recover $f_\pm$ uniquely from $\chi_\gm$. 
In particular,  
  if $f_+$ falls into the null space of $\hat{L}_+$, 
  the information of $f_+$ is lost due to the mapping. 
In this sense, 
  the DC and the SC are not equivalent.

Noticing that the null space of $\hat{L}_+$
  is spanned by $\{ 1, \log\xi_* \}$,  
  we have found from the simulation data
  that $f_+$ indeed behaves like $\ln\xi_*$
  for shorter distances.
The information of this logarithmic behavior in the DC 
  is therefore lost from the SC. 
As a result, 
  the SC has weaker dependence on $\phiA$ than the DC.

Thus we have evidenced 
  that the DC is more informative than the SC. 
The extra information 
  can be obtained from the numerical data 
  by fitting Eq.~(\ref{f+//C+}) 
  to the asymptotic logarithmic behavior of $f_+$.

Interpretation of this information, 
  contained in $C_+$ and $\ld$ 
  obtainable through Eq.~(\ref{f+//C+}),
  is our third question.
It is manifestation of the cage effect 
  at spatiotemporal scales other than usually noticed.
The logarithmic behavior of the DC 
  can be explained 
  by assuming a kind of ``caged'' dynamics for $C_a^\Delta$,  
  such that the cage effect survives for times longer than $\tauA$ 
  at length\-scales greater than $\ell_0$, 
  so that temporal growth of $C_a^\Delta$ is saturated
  as is prescribed in Eq.~(\ref{ve.caged}).
A simple version of such dynamics 
  is exemplified 
  by the elastic modeling in Eq.~(\ref{eqs:dt.psi.muR}).
The dynamics assumed behind Eq.~(\ref{ve.caged}), 
  including the elastic modeling, 
  can account for the logarithmic behavior of $f_+$ 
  and the non-divergent behavior of $f_-$ 
  at the shorter length\-scales. 
Thus the present work 
  contributes to development of solid-based approach to liquid dynamics,
  allowing us to extract 
  experimentally verifiable predictions about DC 
  from theoretical modeling of $C_a^\Delta$, 
  such as the elastic modeling.
The elastic modeling, however, 
  turns out to be too simple 
  to reproduce the $\phiA$-dependence 
  of the coefficient $C_+$ 
  of the logarithmic behavior in Eq.~(\ref{f+//C+}): 
  while the numerical data suggests 
  (at least for $\phiA \le 0.7$)
  that $C_+$ is an increasing function of $\phiA$, 
  the elastic modeling makes the opposite prediction,
  as was noted toward the end of Subsec.~\ref{subsec:log.analysis}.

As the logarithmic behavior of $f_+$ 
  comes from $\hat{L}_+$, which is essentially the 2D Laplacian,
  its counterpart in other dimensionalities 
  (1D and 3D)
  deserves some remarks.
The DC in the 1D setup \cite{Majumdar.PhysicaA177,Ooshida.PRE88}
  is known to behave as 
\begin{equation}
  \Av{{R_i}{R_j}} = 2S\ell_0 \sqrt{\frac{\Dc_*{t}}{\pi}}
  \left( 
  e^{-\vartheta^2} - {\sqrt\pi}\abs{\vartheta} \erfc\abs{\vartheta}
  \right)
  \label{RR.SFD},
\end{equation}
  whose shorter-distance asymptotic form,
  $\Av{{R_i}{R_j}} \propto 1 - {\sqrt\pi}\abs{\vartheta}$,
  is essentially a fundamental solution to the 1D Laplacian.
In the 3D case,
  we expect that the DC in some conditions 
  may behave like the fundamental solution to the 3D Laplacian,
  namely $1/\vartheta$.
Numerical exploration of such behavior 
  would be an interesting direction of investigation.

Let us now remark on what could be done in the near future 
  within the 2D setup.
First of all,
  we would like to invite experimentalists  
  to measure the DC, verify Eq.~(\ref{f+//C+}) 
  and thereby evaluate $C_+$.
Since $C_+$ seems to carry information 
  on some long-lived aspect of the cage effect 
  at length\-scales greater than $\ell_0$, 
  investigation of its dependence on $\tD$ and $\phiA$
  by experiments (both real and simulated) 
  will promote understanding of liquid dynamics at such scales.
On the side of theoreticians,
  it will be desirable
  that they develop more sophisticated modeling 
  for the dynamics of $C_a^\Delta$,
  which, with the aid of the formulae in Subsec.~\ref{subsec:formula}, 
  should allow 
  prediction of $C_+$ as a function of $\phiA$;
  its experimental verification 
  will serve as a touchstone of such modeling.
Viscoelastic modelings of meso\-scopic deformations
  by combination of Maxwell-like phenomenology 
  and the mode-coupling theory (MCT)
  would be a hopeful direction.
While the recent work 
  by Maier \textit{et al.} in this direction 
  is based on Mori--Zwanzig projection formalism 
  of momentum balance \cite{Maier.PRL119},
  we have been developing 
  a different version of MCT 
  for $\Cd^\Delta$ and $\Cr^\Delta$ 
  on the basis of the (over\-damped) Dean--Kawasaki equation 
  \cite{Ooshida.JPS15s,Ooshida.PRE94}.
Details on this version of MCT 
  will be reported elsewhere. 
Here we note, parenthetically,  
  that MCT for tagged particles in Ref.~\cite{Dell.PRE92} 
  predicts DC with spatial oscillation unobserved in the experiment,  
  and also that two completely different ideas 
  for improvement on MCT for a tagged particle 
  are discussed in Refs.~\cite{Abel.PNAS106,Ooshida.PRE88}.

As another possible direction to extend the present work, 
  we may mention cases of higher $\phiA$, 
  which makes the liquid more glassy 
  and the relaxation time long enough 
  to allow comparison 
  between the plateau regime, $\tau_0 \ll \tD \ll \tauA$, 
  and the ``post-$\tauA$'' regime, $\tD \gg \tauA$. 
The elastic model
  is expected to be valid in the plateau regime
  and therefore we may be able to extract 
  information of the elastic moduli 
  from the logarithmic behavior of the DC. 
The ``post-$\tauA$'' regime with long $\tauA$ and high $\phiA$ 
  is where Doliwa and Heuer \cite{Doliwa.PRE61} found 
  the significance of the directional aspect 
  in studying space--time correlations in glassy liquids.
Quantitative analysis of $f_\pm$ in such cases 
  will provide deep insights into caged dynamics, 
  hopefully even more valuable than the impressive pictures 
  of the DC (Fig.~\ref{Fig:RR.2D}) and the SC (Fig.~\ref{Fig:chiGm}).

\begin{acknowledgments}
We express our cordial gratitude 
  to Grzegorz Szamel, Erika Eiser, Peter Schall, 
  Hisao Hayakawa, Norihiro Oyama, Kunimasa Miyazaki, Atsushi Ikeda, 
  Takenobu Nakamura, Hajime Yoshino, Kang Kim, 
  Takahiro Hatano, and So Kitsune\-zaki 
  for fruitful discussions, 
  insightful comments and valuable suggestions.
The first author (Ooshida) also thank 
  Eni Kume and Shiladitya Sengupta 
  for informative discussions 
  on the occasion of recent international conferences 
  held in the difficult time of the Covid19 pandemic.
Encouraging reactions to our arXiv preprint 
  from Laurence Noirez and Alessio Zaccone
  are highly appreciated.
Last but not least, 
  the assistance of Susumu Goto 
  as a coauthor of Ref.~\cite{Ooshida.PRE94}
  is gratefully acknowledged, 
  especially in preparation of Fig.~\ref{Fig:RR.2D}. 
\par
This work was supported 
  by Grants-in-Aid for Scientific Research (\textsc{Kakenhi})  
  JP-15K05213, JP-18K03459 and JP-21K03404, JSPS (Japan).
\end{acknowledgments}

\appendix
\section{Numerical procedure for calculating DC from particle data}
\label{app:DC}

The DC tensor defined by Eq.~(\ref{ChiR}) 
  has essentially only two components, 
  denoted with $\Xl$ and $\Xtr$,
  as is seen in Eq.~(\ref{ChiR.L+T}).
They are computed by extending the numerical method 
  in Appendix~A of Ref.~\cite{Ooshida.PRE88} to the 2D system.
Here we describe 
  some details of formulation and computation.   

Suppose that we have data 
  of $\{ \mb{r}_i(t) \}_{i=1,2,\ldots,N}$, 
  from which we prepare the displacement vector 
  $\mb{R}_i = \mb{R}_i(t,s)$ according to Eq.~(\ref{R=}),  
  and the relative position vector
\[  
  \mb{r}_{ij}(s) = \mb{r}_j(s)-\mb{r}_i(s)
\]
  for all pairs $(i,j)$ within some distance 
  (at most half the system size $L$).
On the basis of these data, 
  the DC tensor is prescribed in Eq.~(\ref{ChiR})
  by means of the conditional average, $\Av{\quad}_{\di}$.

The conditional average 
  is conceptually formulated as 
\begin{equation}
  \Av{O}_{\di}
  = \frac{\sum_{i,j} 
  \Av{O_{ij} \delta(\mb{r}_{ij}(s) - \di)}%
  }{\sum_{i,j} 
  \Av{\delta(\mb{r}_{ij}(s) - \di)}
  }
  \label{avr2p.observable}
\end{equation}
  for any physical observable $O_{ij}$ 
  associated with the particle pair $(i,j)$. 
In the case of the DCs,
  as the observable $O_{ij}$ we take 
\begin{align}
  (\Xl)_{ij} &= 
  ({\mathbf{e}_{ij}}\cdot\mb{R}_i)
  ({\mathbf{e}_{ij}}\cdot\mb{R}_j)  \relax,
  \\
  (\Xtr)_{ij} &= 
  \det(\mathbf{e}_{ij}, {\mb{R}_i})
  \det(\mathbf{e}_{ij}, {\mb{R}_j}) \relax,
\end{align}
  using orthogonal decomposition of the displacement vectors 
  with regard to 
\begin{equation}
  \mathbf{e}_{ij} 
  = \frac{\mb{r}_{ij}(s)}{\abs{\mb{r}_{ij}(s)}}
  \relax.
\end{equation} 

To evaluate 
  $\Xl (\di[\relax],\tD) = \Av{\Xl }_{\di}$ and  
  $\Xtr(\di[\relax],\tD) = \Av{\Xtr}_{\di}$ numerically, 
  we approximate the delta function in Eq.~(\ref{avr2p.observable}) 
  by a statistical bin with $\Delta{r}$ in width
  (we use $\Delta{r} = L/200 \approx \sigma/3$
  for the present calculations with $N = 4000$), 
  assuming the statistical isotropy of the system at once.
Thus Eq.~(\ref{avr2p.observable}) is discretized 
  as 
\begin{equation}
  \Av{O}_{\di}
  = \frac{%
  \Av{\sum_{ij} O_{i,j} 
  \Theta(\di[\relax] \le \abs{\mb{r}_{ij}(s)} < \di[\relax]+\Delta{r})}%
  }{%
  \Av{\sum_{i,j} 
  \Theta(\di[\relax] \le \abs{\mb{r}_{ij}(s)} < \di[\relax]+\Delta{r})}%
  }%
  \label{bin.observable}, 
\end{equation}
  where $\Theta$ denotes 
  the indicator function of the statistical bin, 
  such that its value equals unity 
  if and only if $\abs{\mb{r}_{ij}(s)}$ satisfies the inequality 
  and otherwise zero.
Finally, with the ensemble average 
  that remains in Eq.~(\ref{bin.observable}) 
  taken over many runs, 
  we obtain $\Xl$ and $\Xtr$ numerically.

\section{Wavenumber integrals involving Bessel function and Gaussian}
\label{app:intK.Bessel.gauss}

For analytical calculations,
  we define 
\begin{equation}
  I_{m,n}(\xi,\tau) 
  = \int_0^\infty e^{-{\tau}k^2 } J_m(k\xi)\,k^{n-1}\D{k}
  \label{int.mn}
\end{equation}
  where $(m,n)$ is a pair of non-negative integers,
  $\xi$ and $\tau$ are positive real numbers, 
  and $J_m$ denotes the Bessel function of order $m$.
Evaluation of this integral 
  is required in Subsec.~\ref{subsec:elastic}
  for $(m,n) = (0,2)$, $(2,2)$ and $(4,2)$.

Let us begin 
  with $I_{0,2}(\xi,\tau)$.
Making use of well-known relations between $J_0$ and $J_1$
  \cite{Arfken.Book2013}, 
  we find 
\begin{gather}
  2\tau I_{0,2}(\xi,\tau) = 1 - \xi I_{1,1}(\xi,\tau)   
  \label{I02//I11},\\
  \dd{}{\xi}\left[ \xi I_{1,1}(\xi,\tau) \right] 
  = \xi I_{0,2}(\xi,\tau)
  \label{Dx.I11//I02},
\end{gather}
  which yields
\begin{equation}
  2\tau \dd{I_{0,2}(\xi,\tau)}{\xi} = -{\xi} I_{0,2}(\xi,\tau)
\end{equation}
  upon elimination of $I_{1,1}(\xi,\tau)$.
This is readily integrated
  to give 
\begin{equation}
  I_{0,2}(\xi,\tau) 
  = \frac{1}{2\tau} \exp\left({-\frac{\xi^2}{4\tau}}\right)
  \label{I02=}
\end{equation}
  with the initial condition, $I_{0,2}(0,\tau) = 1/(2\tau)$, 
  taken into account.  

Subsequently, 
  in order to evaluate $I_{2,2}(\xi,\tau)$, 
  we notice that the derivative of $J_1$ can be expressed 
  in terms of $J_0$ and $J_2$, 
  which allows us to find 
\begin{equation}
  2\dd{I_{1,1}(\xi,\tau)}{\xi} 
  =  I_{0,2}(\xi,\tau) - I_{2,2}(\xi,\tau)
  \label{2Dx.I11}.
\end{equation}
Combining Eq.~(\ref{2Dx.I11}) with Eq.~(\ref{I02//I11}), 
  after some calculation, 
  we have 
\begin{equation}
  I_{2,2}(\xi,\tau) 
  = {\frac{2}{\xi^2}}
  - \frac{1}{2\tau}\left( 1 + \frac{4\tau}{\xi^2} \right) 
  \exp\left( -\frac{\xi^2}{4\tau} \right) 
  \label{I22=}.
\end{equation}
In an analogous way, we also find 
\begin{multline}
  I_{4,2}(\xi,\tau) 
  = 
  \frac{4}{\xi^2}\left( 1 - \frac{12\tau}{\xi^2} \right)
  \brX {} +
  \frac{1}{2\tau}\left(
  1 + \frac{16\tau}{\xi^2} + \frac{96\tau^2}{\xi^4} 
  \right)
  \exp\left( -\frac{\xi^2}{4\tau} \right)   
  \label{I42=},\hFil 
\end{multline}
  making further usage 
  of the recurrence relations for the Bessel function.


\begin{thebibliography}{48}
\bibitem [{not({\natexlab{a}})}]{note.Av}%
  \BibitemOpen
  \bibinfo {note} {By $\Av{\quad}$ we denote suitable ensemble averaging for
  the equilibrium state under consideration. In particular, in the context of
  Langevin dynamics, it denotes averaging over different noise realizations and
  initial conditions. Conditional averaging is denoted with a subscript, as in
  $\Av{\quad}_{\di}$.}\BibitemShut {Stop}%
\bibitem [{not({\natexlab{b}})}]{note.deltaE}%
  \BibitemOpen
  \bibinfo {note} {We adopt $\de(\mb{r}) = (2\pi\varepsilon^2)^{-1}
  \exp(-\mb{r}^2/(2\varepsilon^2))$, with $\varepsilon$ taken to be some
  microscopic length scale. Specifically, we use $\varepsilon = \sigma$ for the
  numerical calculations in Sec.~\protect\ref{sec:num}.}\BibitemShut {Stop}%
\bibitem [{not({\natexlab{d}})}]{note.muR}%
  \BibitemOpen
  \bibinfo {note} {The definition of $\muR$ in the present article is slightly
  different from its counterpart in the reference: $\muR{\Dc_*} =
  (\muR/S){D_*}$ in Eq.~(\ref{Cr.Delta.exp}) corresponds to $\muR{D_*}$ in
  Eq.~(6.11) of Ref.~\cite{Ooshida.PRE94}.}\BibitemShut {Stop}%
\end{thebibliography}
%

\end{document}